\documentstyle{article}
\def\ftoday{{\sl  \number\day \space\ifcase\month 
\or Janvier\or F\'evrier\or Mars\or avril\or Mai
\or Juin\or Juillet\or Ao\^ut\or Septembre\or Octobre
\or Novembre \or D\'ecembre\fi
\space  \number\year}}    
\newcommand{\journal}[4]{{\em #1~}#2\,(19#3)\,#4;}

\newcommand{\hpa}{\journal {Helv. Phys. Acta}}

\newcommand{\ijmp}{\journal {Int. J. Mod. Phys.}}
\newcommand{\ijtp}{\journal {Int. J. Theor. Phys.}}

\newcommand{\pr}{\journal {Phys. Rev.}}

\newcommand{\jp}{\journal {J. Phys.}}

\newcommand{\cmp}{\journal {Commun. Math. Phys.}}
\newcommand{\ctp}{\journal {Commun. Theor. Phys.}}
\newcommand{\cqg}{\journal {Class. Quantum Grav.}}

\newcommand{\np}{\journal {Nucl. Phys.}}
\newcommand{\pl}{\journal {Phys. Lett.}}
\newcommand{\mpl}{\journal {Mod. Phys. Lett.}}
\newcommand{\prep}{\journal {Phys. Rep.}}

\makeatletter
\@addtoreset{equation}{section}
\makeatother

\newcommand{\es}{\\[3mm]}

\renewcommand{\a}{\alpha}
\renewcommand{\b}{\beta}
\newcommand{\g}{\gamma}           \newcommand{\G}{\Gamma}
\renewcommand{\d}{\delta}         \newcommand{\D}{\Delta}
\newcommand{\e}{\varepsilon}
\newcommand{\k}{\kappa}
\newcommand{\la}{\lambda}        \newcommand{\LA}{\Lambda}
\newcommand{\m}{\mu}
\newcommand{\n}{\nu}
\newcommand{\om}{\omega}         \newcommand{\OM}{\Omega}
\newcommand{\p}{\psi}              
\newcommand{\r}{\rho}
\newcommand{\s}{\sigma}           \renewcommand{\S}{\Sigma}
\newcommand{\th}{\theta}         
\newcommand{\f}{{\phi}}           \newcommand{\F}{{\Phi}}
\newcommand{\vf}{{\varphi}}

\renewcommand{\AA}{{\cal A}}
\newcommand{\BB}{{\cal B}}

\newcommand{\DD}{{\cal D}}

\newcommand{\FF}{{\cal F}}
\newcommand{\GG}{{\cal G}}

\newcommand{\LL}{{\cal L}}

\newcommand{\NN}{{\cal N}}

\newcommand{\SS}{{\cal S}}

\newcommand{\xint}{\dint d^4x\;}
\newcommand{\sla}{\raise.15ex\hbox{$/$}\kern -.57em} 
\newcommand{\Sla}{\raise.15ex\hbox{$/$}\kern -.70em}
\def\h{\hbar}
\def\Lp{\displaystyle{\biggl(}}
\def\Rp{\displaystyle{\biggr)}}
\def\LP{\displaystyle{\Biggl(}}

\newcommand{\lp}{\left(}\newcommand{\rp}{\right)}
\newcommand{\lc}{\left[}\newcommand{\rc}{\right]}
\newcommand{\lac}{\left\{}\newcommand{\rac}{\right\}}

\newcommand{\complex}{{\kern .1em {\raise .47ex
\hbox {$\scriptscriptstyle |$}}
    \kern -.4em {\rm C}}}
\newcommand{\real}{{{\rm I} \kern -.19em {\rm R}}}
\newcommand{\rational}{{\kern .1em {\raise .47ex
\hbox{$\scripscriptstyle |$}}
    \kern -.35em {\rm Q}}}
\renewcommand{\natural}{{\vrule height 1.6ex width
.05em depth 0ex \kern -.35em {\rm N}}}

\newcommand{\tr}{{\rm {Tr} \,}}
\newcommand{\half}{\dfrac{1}{2}}
\newcommand{\pa}{\partial}

\newcommand{\fud}[2]{{\frac{\delta #1}{\delta #2}}}
\newcommand{\dpad}[2]{{\displaystyle{\frac{\partial #1}{\partial #2}}}}
\newcommand{\dfud}[2]{{\displaystyle{\frac{\delta #1}{\delta #2}}}}
\newcommand{\dfrac}[2]{{\displaystyle{\frac{#1}{#2}}}}
\newcommand{\dsum}[2]{\displaystyle{\sum_{#1}^{#2}}}   
\newcommand{\dint}{\displaystyle{\int}}

\newcommand{\twiddle}{\lower.9ex\rlap{$\kern -.1em\scriptstyle\sim$}}

\newcommand{\vev}[1]{\left\langle {#1}\right\rangle}
\newcommand{\equ}[1]{(\ref{#1})}
\newcommand{\eq}{\begin{equation}}
\newcommand{\eqn}[1]{\label{#1}\end{equation}}
\newcommand{\eea}{\end{eqnarray}}
\newcommand{\eqa}{\begin{eqnarray}}
\newcommand{\eqan}[1]{\label{#1}\end{eqnarray}}
\newcommand{\ba}{\begin{array}}
\newcommand{\ea}{\end{array}}
\newcommand{\eqac}{\begin{equation}\begin{array}{rcl}}
\newcommand{\eqacn}[1]{\end{array}\label{#1}\end{equation}}

\newcommand{\point}[1]{\vspace{3mm}

\noindent{\bf #1}}
\newcommand{\remark}{\vspace{3mm}

\noindent{\large{\bf Remark. }}}
\newcommand{\kramer}{\vspace{2mm}

\noindent}
\newcommand{\remarks}{\vspace{3mm}

\noindent{\large{\bf Remarks.}}\begin{enumerate} }
\newcommand{\skramer}{\end{enumerate}

\noindent}
\newcommand{\bi}{\begin{itemize}} \newcommand{\ei}{\end{itemize}}
\newcommand{\pint}{\dint d^4p\;}

\newcommand{\bint}[2]{{\dint_{\kern -.4em #1}^{#2}}}
\newcommand{\ZC}{Z^{\rm c}}

\newcommand{\Bsurl}[1]{ 
{\raise 5ex \hbox{$\overline{{\raise -5ex \hbox{$#1$} }}$} }
}
\newcommand{\bsurl}[1]{ 
{\raise 3ex \hbox{$\overline{{\raise -3ex \hbox{$#1$} }}$} }
}
\newcommand{\surl}[1]{ 
{\raise 2ex \hbox{$\overline{{\raise -2ex \hbox{$#1$} }}$} }
}
\newcommand{\iv}{\dint dV\;}
\newcommand{\is}{\dint dS\;} \newcommand{\isb}{\dint d\bar S\;}
\newcommand{\ad}{{\dot\a}}  \newcommand{\bd}{{\dot\b}}  
\newcommand{\gd}{{\dot\g}}  \newcommand{\dd}{{\dot\d}}  
  
\newcommand{\QB}{{\bar{Q}}}    
\newcommand{\DB}{{\bar{D}}}
\newcommand{\BBAR}{{\bar{B}}}
\newcommand{\SB}{{\bar{S}}}
\newcommand{\AB}{{\bar{A}}}
\newcommand{\JB}{{\bar{J}}}
\newcommand{\cb}{{\bar{c}}}
\newcommand{\db}{{\bar\d}}
\renewcommand{\sb}{{\bar\s}}
\newcommand{\tb}{{\bar\th}}
\newcommand{\psb}{{\bar\p}} 
\newcommand{\chib}{{\bar\chi}}
\newcommand{\smuaad}{\s^\m_{\a\ad}}
\newcommand{\sbmuaad}{{\bar\s}_\m^{\ad\a}}

\begin{document}
\thispagestyle{empty}
\begin{center}
{\Huge {\bf Supersymmetry, Supercurrent,\\[1mm] and Scale Invariance}}
\vspace{8mm}

{\Large Olivier Piguet}\footnote{On leave of
absence from:
{\it D\'epartement de Physique Th\'eorique --
     Universit\'e de Gen\`eve, 24 quai E. Ansermet -- CH-1211 Gen\`eve
     4, Switzerland.}}$^,$\footnote{Supported in part by the 
     Swiss National Science Foundation and the Brazilian National Research 
     Council (CNPq)}
\vspace{3mm}

{\it Instituto da F{\'\i}sica, Universidade Cat\'olica de Petr\'opolis\\
25610-130 Petr\'opolis, RJ, Brazil\\ \vspace{.5mm}

and\\ \vspace{.5mm}

Centro Brasileiro de Pesquisas F{\'\i}sicas (CBPF)\\
Rua Xavier Sigaud 150, 22290-180 Urca, RJ, Brazil}
\vspace{10mm}

CBPF--NF--072/96\\
UGVA--DPT 1996/08--938\\
hep-th/9611003
\vspace{1cm}

{\large Lectures given at the Catholic University of Petr\'opolis (RJ) and at
the CBPF (Rio de Janeiro), September 1995 and July-August 1996\\
(Notes written in collaboration with Oswaldo Monteiro Del Cima)}
\vspace{1cm}

\end{center}
\vspace{5mm}

\tableofcontents
\newpage

\section*{Preface}
These notes are an expanded version of a set of lectures I have given at
the Catholic University of Petr\'opolis (UCP) and at the Centro 
Brasileiro de Pesquisas F{\'\i}sicas (CBPF).

I thank all the members of the theoretical physics departments 
of both  institutions for their very warm hospitality, and the students
who assisted to these lectures, with the hope that they have taken
profit from them.

I feel me particularly indebted towards my friends Renato M. Doria
(UCP), Jos\'e A. Helayel Neto, Marco Antonio de Andrade, 
Oswaldo M. Del Cima (CBPF) and Silvio P. Sorella (UERJ)
 for their continuous assistance during my stays in Brazil and
for the many interesting scientific discussions we had together.

I must specially thank Oswaldo M. Del Cima for his very valuable help
for writing these notes.

Finally I thank the Brazilian National Research Council (CNPq) and the CBPF
for their generous financial assistance.

\section{Introduction}\label{introduction}
The aim of the present lectures is to give an introduction to the
renormalization of supersymmetric gauge theories in 4-dimensional
space-time. This will include the analysis of the ultraviolet divergences,
and much emphasis will be put on the so-called ``ultraviolet finite''
models. Exemples of the latters might be relevant as realistic ``grand
unified theories'' of the particle interactions.

Some ``textbook knowledge'' of renormalization theory 
is expected from the
listeners. The approach I shall follow is that of ``algebraic
renormalization'', see e.g.~\cite{pigsor-book}. 
On the other hand, the supersymmetry
formalism, in particular the superspace formalism developed
in these lectures, is not supposed to be known in advance. One may however
consult the classical textbooks on the
subject~\cite{gates-book,bagger-book,west-book,buchbinder}, as well as
reviews such as the ones collected in~\cite{reviews}. 
The book~\cite{ps-book} also
presents this formalism, with more emphasis on the problem 
of renormalization. I shall follow the notations and conventions of
~\cite{ps-book}.  

Usual symmetries, either of the space-time type -- e.g. Poincar\'e -- or
of the internal type -- e.g. 
U(1)$\times$SU(2)$\times$SU(3) or SU(5) -- are described by Lie 
groups~\cite{group-th}.
Is it possible to unify  both types of supersymmetry?
The ``no-go theorem'' of Coleman and Mandula~\cite{col-mand}
answers by the negative. More precisely, it states that any Lie group
containing the Poincar\'e group and an internal symmetry group as
maximal subgroups is the trivial product of both. 
In other words, internal
symmetry transformations always commute with the Poincar\'e
transformations. 

The hypotheses of this theorem are quite general. 
They consist in the axioms of 
relativistic quantum field theory~\cite{haag}, in the existence of a
unitary $S$-matrix and in the assumption that all symmetries are realised
in terms of Lie groups. 
A way to circumvent it was however found by Haag, {\L}opuschanski and
Sohnius~\cite{haag-lo-so}. These authors simply relaxed one of the
hypotheses of the no-go theorem, namely the one which concerns the
groups of symmetry. They assumed that the 
infinitesimal generators of the symmetry  obey a {\it superalgebra}.
A superalgebra is a generalization of the notion of a Lie algebra, where
some of the infinitesimal generators are fermionic, which means that
some of the commutation rules are replaced by anticommutation rules.
The result of~\cite{haag-lo-so} is still very restrictive: the only
superalgebras compatible with the general axioms of
relativistic quantum  field theory and with the unitarity of the $S$-matrix
are the supersymmetries of the Wess-Zumino type, i.e. those where the
fermionic  generators carry a spin $1/2$.

Another theoretical motivation for studying supersymmetry is offered by
{\it string theory}~\cite{gr-sch-witten}. Indeed, the presence of
fermionic string states together with 
bosonic ones, imposes a supersymmetric structure to the theory.
In the effective
field theories which approximate string theory in the  
energy domain  below
the Planck mass, equal to $10^{19}$~GeV, this structure
manifests itself as a Wess-Zumino supersymmetry.

A further motivation for supersymmetry is found in the solution of the 
{\it hierarchy problem}~\cite{hierarchy,girardello-g} 
of the grand unified theories.
In these theories~\cite{GUTs}, 
which tend to unify all the particles 
and forces described by the standard model of particle 
interactions~\cite{standard-mod},
two energy scales must be introduced, typically of the order of $10^3$
- $10^4$ GeV -- the electro-weak scale --
and $10^{15}$ - $10^{16}$ GeV -- the grand unification scale. 
This means that one has to ``fine tune'' a mass difference
expressed by a number with more than 12 significative digits! 
This fine tuning would be perfectly utopic in the framework of
conventional gauge theories, since the presence of quadratic divergences
of the mass corrections induces a strong instability of the  difference
of the renormalized masses, which must be fine tuned at each order of the
perturbative calculus. The interest in considering  supersymmetric
theories is that ultraviolet divergences are milder, in particular
the mass corrections depend only on the logarithm of the ultraviolet
cut-off, instead  of its square. The huge mass differences in grand
unified theories are then much more stable\footnote{Such a picture is
more understandable, in physical terms, 
within a framework where one considers the field
theoretical model as an effective field theory, the ultraviolet cut-off
being a physical parameter of an hypothetical exact theory 
-- e.g. string theory -- describing
the phenomena at very high energies. 
This parameter might be the Planck mass.}.

Supersymmetry having thus a tendency to cancel some of
the ultraviolet divergences,
a natural question to ask is: could supersymmetry 
eventually lead to a
complete cancellation of these divergences?
Let us mention that searches for general ultraviolet  
finite models have been done -- up to the order of the two-loop graphs.
They have lead to the conclusion that supersymmetry is most likely 
required~\cite{lucha-neufeld,bohm-denner}. 

Some ultraviolet finite supersymmetric models
have been known since a long time.
All these
models had an extended supersymmetry: $N$=4~\cite{mandelstam} or 
$N$=2~\cite{howe-s-w,howe-s-t}, where $N$ counts the fermionic 
generators. 
However, gauge models with extended supersymmetry are 
not physically appealing since
they don't accomodate chiral fermions --
in contrast with the $N$=1 models. More recently, 
finite models with $N$=1 supersymmetry were proposed.  
A complete list of such models, finite at least up to 
the two-loop order~\cite{par-west},
was first obtained in~\cite{hamidi-p-s,rajpoot-t}. Then 
some 
proposals
for all order finiteness were done~\cite{jones-kazakov,chinois}. 
A common
feature of these finite $N$=1 supersymmetric models is that 
they are based on a simple gauge group -- hence they possess a single gauge
coupling constant -- and also that 
their Yukawa coupling constants must be functions 
of the gauge coupling constant. 
This indicates them as valuable candidates for
grand unified theories, which moreover possess
the power to predict the fermion
masses since the Yukawa couplings
are no more arbitrary parameters, in contrast to
the usual, i.e. nonfinite, grand unifications.

Finally, a general criterion for the all order finiteness was 
given~\cite{ps-fin-pl}-\cite{alushta}. 
This criterion states a set of
necessary and sufficient conditions for a theory to have of all its
Callan-Symanzik ``$\beta$-functions'' vanishing 
to all orders of perturbation theory. Only
the knowledge of the general expression for
the one loop $\beta$-functions~\cite{gross-wil} is required.
The physical meaning of vanishing $\b$
is the absence of scale anomalies,
hence the scale invariance of the theory -- at least asymptotically if massive
particles are present. This does not mean complete ultraviolet
finiteness, since infinite renormalizations of the field amplitudes are
still allowed. The nonphysical character of the 
latter~\cite{pigsor-book} however 
justifies the terminology of ``ultraviolet finiteness''. 

Applications of the criterion of ultraviolet finiteness
to realistic models based on the grand
unification group SU(5) with three fermion generations have been
performed recently~\cite{zoup,claudio} (see also~\cite{jones-kazakov}
for a different approach.)

\newpage
\section{Generalities}\label{generalites}
\subsection{Extended Supersymmetry Algebra}\label{algebre susy}
The basis of the extended $N$-supersymmetry algebra 
consists of~\cite{haag-lo-so}:
\bi
\item
Bosonic ({\it even}) hermitean generators $T_a$, 
$a=1\cdots \mbox{dim}(G)$, of some Lie group $G$,  
\item The generators $P_\m$ and $M_{[\m\n]}$ of the 4-dimensional
Poincar\'e group.
\item
Fermionic ({\it odd}) generators $Q_\a^i$, 
$\a=1,2$; $i=1,\cdots,N$ belonging to a dimension $N$ representation 
of $G$, 
and  their conjugates $\QB^\ad_i$. 
\item 
Central charges $Z^{[ij]}$, i.e. bosonic operators commuting with all the
$T_a$'s and all the $Q_\a$'s and $\QB^\ad$'s, as well as with the
Poincar\'e generators.
\ei
The $T_a$'s and $Z^{[ij]}$'s are scalars, whereas the $Q_\a^i$'s belong to the
representation $(1/2,0)$ of the Lorentz group and the $\QB^\ad_i$'s to
the conjugate representation $(0,1/2)$. The latters are written as 
Weyl spinors, with two complex 
components\footnote{The notations and conventions are
detailed in Appendix~\ref{notations}.}.  

The general  superalgebra of $N$-extended supersymmetry, also
called the $N$-super-Poincar\'e algebra, reads
(we write only the nonvanishing (anti)commutators):
\eq\ba{l}
[M_{\m\n},M_{\r\s}] = -i(g_{\m\r}M_{\n\s} - g_{\m\s}M_{\n\r} +
                        g_{\n\s}M_{\m\r} - g_{\n\r}M_{\m\s} )\ ,\es
[M_{\m\n},P_\la] = i(P_\m g_{\n\la} - P_\n g_{\m\la})\ ,
\ea\eqn{alg-poincare}
\eq\ba{l}
[T_a,T_b] = i {f_{ab}}^c T_c\ ,
\ea\eqn{alg-lie}
\eq\ba{l}
\lac Q^i_\a,Q^j_\b \rac = \e_{\a\b} Z^{[ij]}\ ,\es
\lac Q^i_\a,\QB^j_\ad \rac = 2\d^{ij} \smuaad P_\m  \ ,\
\ea\eqn{alg-susy}
\eq\ba{l}
[Q^i_\a,M_{\m\n}]= \dfrac{1}{2} {(\s_{\m\n})_\a}^\b Q^i_\b\ ,\es
[Q^i_\a,T_a] = {(R_a)^i}_j Q^j_\a\ . 
\ea\eqn{alg-lie-susy}
This result is the most general one for a massive theory. 
In a massless theory, another
set of fermionic charges, $S^i_\a$ (and their conjugates), 
may be present. 
Then, the Lie group $G$ is U($N$) for $N\not=4$, and
either U(4) or SU(4) for $N$=4. The superalgebra moreover 
contains all the
generators of the conformal group -- which contains the Poincar\'e group as
a subgroup: one calls it the $N$-superconformal algebra.

In these lectures we will restrict ourselves to the case $N$=1.
\subsection{$N$=1 Superfields}
In the $N$=1 case, the part of the superalgebra 
\equ{alg-poincare}-\equ{alg-lie-susy} 
which involves the spinor charges reduces 
to the original Wess-Zumino algebra 
\eq
\lac Q_\a,\QB_\ad \rac = 2 \smuaad P_\m  \ ,\quad
\lac Q_\a,Q_\b \rac = 0\ ,\quad \lac \QB_\ad,\QB_\bd \rac = 0\ ,
\eqn{alg-wz}
to
\eq\ba{l}
[Q_\a,M_{\m\n}]= \dfrac{1}{2} {(\s_{\m\n})_\a}^\b Q_\b\ ,\quad
[\QB^\ad,M_{\m\n}]= -\dfrac{1}{2} {(\sb_{\m\n})^\ad}_\bd \QB^\bd\ ,\es
[Q_\a,P_\m]= 0\ ,\quad [\QB_\ad,P_\m]= 0\ ,
\ea\eqn{com-poinc-susy}
and to
\eq
[ Q_\a,R] = -Q_\a\ ,\quad [\QB_\ad,R] = \QB_\ad\ .
\eqn{com-r-susy} 
Here, $R$ is the infinitesimal generator of an Abelian group into
which the internal symmetry group $G$ has shrunk.

The objects which transform covariantly under the supersymmetry
transformations  are the superfields\footnote{See  
Appendix~\ref{notations} for
the definitions, notations and conventions.}, 
either of the general type, or of the chiral type. As
explained in Appendix~\ref{notations}, 
a superfield is a superspace function $\f(x^\m,\th_\a,\tb_\ad)$, where 
$\th_\a$, $\a=1,2$, are complex Grassmann variables, and $\tb_\ad$ their
complex conjugates.
A chiral superfield $A(x,\th,\tb)$, resp. antichiral 
superfield $\bar A(x,\th,\tb)$, is a superfield obeying
the constraint 
\eq
\DB_\ad A = 0\ ,\quad\mbox{resp.}\quad D_\a \bar A = 0\ ,
\eqn{chiral-const}
where $D_\a$, $\DB_\ad$ are the covariant superspace 
derivatives \equ{def-cov-der}.
The component fields of a superfield span a supermultiplet, i.e. an
irreducible representation of the supersymmetry algebra.
The translation, supersymmetry and $R$ transformation 
laws of a superfield $\f$
are defined by the superspace differential operators
\eq\ba{l}
\d^P_\m \f = \pa_\m\f\ ,\es
\d^Q_\a \f = \lp \dpad{}{\th^\a} + i\smuaad\tb^\ad\pa_\m\rp \f , \es
\d^{\bar Q}_\ad \f = \lp -\dpad{}{\tb^\ad} - i\th^\a\smuaad\pa_\m\rp \f\ ,
\ea\eqn{delta-susy}
and
\eq
\d^R\f = i\lp n+\th^\a\dpad{}{\th^\a} - \tb^\ad\dpad{}{\tb^\ad}\rp \f\ .
\eqn{r-transf}
In the last equation the real number $n$ is the 
``$R$-weight'' of the superfield $\f$.
The $R$-weigths of a pair of complex conjugates superfields are opposite
to each other.
These differential operators fulfil the algebra
\eq\ba{l}
\lac \d^Q_\a,\d^{\bar Q}_\ad \rac = -2i\smuaad\d^P_\m\ ,\es
[ \d^Q_\a,\d^R ] = i\d^Q_\a\ ,\quad
[ \d^\QB_\ad,\d^R ] = -i\d^\QB_\ad\ ,\es
\mbox{(the other (anti)commutators vanishing)}\ .
\ea\eqn{alg-diff-op}

\subsection{Invariant Actions and Ward Identity Operators}
A supersymmetric classical action $\S$ is given by the superspace integral 
-- as defined by \equ{integr} -- of
some local functional of the superfields entering the considered theory,
and of their covariant derivatives. Such integrals are indeed
invariant under supersymmetry transformations.

The actions which will be considered in these lectures will be invariant
as well under other symmetry transformations. These invariances will be
expressed in a functional way. Let denote by $\d_X\vf$ the infinitesimal 
transformation of the superfield $\vf$ along the generator $X$ of the
(super)group of symmetries, e.g. one of the transformations 
\equ{delta-susy}, \equ{r-transf}. Let us define the associated functional
Ward identity (WI) operator as the differential operator
\eq
W_X: = -i \dsum{\vf}{} \dint \d_X\vf\dfud{}{\vf}\ .
\eqn{wi-oper}
The summation runs over all superfields $\vf$.
The superspace functional derivatives are defined by \equ{funct-diff}.
We don't specify the integration measure, which is $dV$, $dS$ or 
$d\SB$ according to the type of $\vf$.

The invariance of the classical action $\S$ is then expressed by the 
{\it Ward identity} (WI) 
\eq
W_X\S=0\ .
\eqn{ward-id}

An important property of the the WI operators is that they obey the
superalgebra 
\eq
\lc W_{X_a},W_{X_b}\rc = i f_{abc} W_{X_c}\ ,
\eqn{wi-op-alg}
if the differential operators or matrices $\d_{X}$ obey the
(anti)commutation rules
\eq
\lc \d_{X_a},\d_{X_b}\rc =  f_{abc} \d_{X_c}\ .
\eqn{diff-op-alg}
In the equations above the brackets are ``graded commutators'', i.e.
anticommutators $\{\,,\,\}$ if both arguments are odd, and 
commutators $[\,,\,]$ otherwise.

As a rule, the WI operators obey the same (super)algebra as the abstract
(super)algebra of the generators, the superalgebra \equ{alg-wz} - 
\equ{com-r-susy} for instance.

\newpage
\section{The Baby Model}
\subsection{The Action}
The simplest $N$=1 supersymmetric model in four dimensions is the model
of Wess and Zumino\cite{w-z-baby}, which consists of a chiral superfield
$A$ in self-interaction. Its action reads
\eq
\S = \dfrac{1}{16}\iv A \bar A   + \is W(A) + \isb \bar W(\bar A)\ ,
\eqn{baby-action}
with the {\it superpotential}
\eq
W(A) = \dfrac{1}{4}\lp \dfrac{m}{2}A^2 +\dfrac{\la}{12}A^3 \rp\ ,
\eqn{baby-s-pot}
the mass $m$ and coupling constant $\la$ being real.
In components \equ{comp-chiral}, we have
\eq\ba{l}
\S= \xint \lp F\bar F +\dfrac{i}{2}\p\s^\m\pa_\m\psb + 
    \pa^\m A\pa_\m\bar A     \right.\es\phantom{S= }\left.
  -\dfrac{m}{4}(4AF-\p^2 + \mbox{conj.})
  -\dfrac{\la}{8}(2A^2F-A\p^2+\mbox{conj.}) \rp
\ea\eqn{baby-comp-action}
One sees that the complex scalar field $F$ is auxiliary, i.e.
its equation of motion can be solved algebraically:
\eq
F=F(\bar A) = 4{\bar W}'(\bar A) = mA + \dfrac{\la}{4}A^2\ .
\eqn{baby-f-eq}
We may, if we want, insert this into the action, obtaining 
\[
\S=\xint\lp \dfrac{i}{2}\p\s^\m\pa_\m\psb + 
\pa^\m A\pa_\m\bar A 
 +\dfrac{m}{4}\p^2+\dfrac{m}{4}{\psb}^2+
\dfrac{\la}{8}A\p^2+\dfrac{\la}{8}{\bar A}\psb^2 -
V(A,\bar A)\rp \ ,
\]
with a potential given by
\[
V(A,\bar A) = F(\bar A)\bar F(A) 
  = \left\vert mA+\dfrac{\la}{4}A^2\right\vert^2   \ ,
\]
which turns out to be positive.
\subsection{Field Equations}
The field equations read
\eq\ba{l}
\dfud{\S}{A} =\dfrac{1}{16} \DB^2\AB +\dfrac{m}{4}A+
       \dfrac{\la}{16} A^2 = 0\es
\dfud{\S}{\AB} = \dfrac{1}{16} D^2A + \dfrac{m}{4}\AB+
      \dfrac{\la}{16}\AB^2 = 0\ .
\ea\eqn{baby-eq}
One may combine them in order to find
\eq\ba{l}
4m\dfud{\S}{A} -\DB^2\dfud{\S}{\AB} =
               (\pa^2+m^2)A + \mbox{interaction} = 0\es
4m\dfud{\S}{\AB} -D^2\dfud{\S}{A} =
               (\pa^2+m^2)\AB + \mbox{interaction} = 0\ .
\ea\eqn{baby-k-g}
\subsection{Free Propagators}
The computation of the free propagators amounts to compute
the Green functions of the theory without self-interaction -- i.e. with
$\la=0$ -- but in presence of an external chiral superfield
source  $J$ coupled to $A$. This is described by the action
\eq
\S_J = \dfrac{1}{16}\iv A \bar A   + \dfrac{m}{8}\is A^2 + 
 \dfrac{m}{8}\isb \AB^2 +\is JA + \isb\JB\AB\ ,
\eqn{baby-action-j}
leading to the field equations
\eq\ba{l}
\dfrac{1}{16} \DB^2\AB +\dfrac{m}{4}A = -J\es
\dfrac{1}{16} D^2A + \dfrac{m}{4}\AB = -\JB\ .
\ea\eqn{baby-eq-j}
Combining them as above, we find
\eq\ba{l}
(\pa^2+m^2)A  = \DB^2\JB-4mJ \es
(\pa^2+m^2)\AB  = D^2J-4m\JB \ ,
\ea\eqn{baby-k-g-j}
The equations for the propagators are obtained by differentiating with
respect to the sources:
\eq\ba{l}
(\pa^2+m^2)\D_{AA}(1,2)  = (\pa^2+m^2)\dfrac{\d A(1)}{i\d J(2)} 
  = 4im \d_S(1,2) \ ,\es
(\pa^2+m^2)\D_{A\AB}(1,2)  = (\pa^2+m^2)\dfrac{\d A(1)}{i\d\JB(2)} 
  = -i\DB^2 \d_\SB(1,2) \ ,\es
(\pa^2+m^2)\D_{\AB\AB}(1,2)  = (\pa^2+m^2)\dfrac{\d\AB(1)}{i\d\JB(2)} 
  = 4im \d_\SB(1,2) \ ,
\ea\eqn{baby-green-eq}
where $\d_S$ and $\d_\SB$ are the chiral and antichiral  superspace Dirac
distributions given by \equ{delta-funct}. The notation $f(1)$ means
$f(x_1,\th_1,\tb_1)$, etc. 

In order to solve the latter system, one introduces 
the causal scalar propagator
$\D_c(x)$ defined as a particular inverse of the Klein-Gordon operator:
\eq\ba{l}
i(\pa^2+m^2)\D_c(x) = \d^4(x)\ :\es
\D_c(x) = \dfrac{1}{(2\pi)^4}\pint e^{ipx}\tilde \D_c(p)\ ,\quad
\tilde\D_c(p) = \dfrac{i}{p^2-m^2+i0}\ ,
\ea\eqn{scalar-prop}
with the notation
\[
\dfrac{1}{z+i0}:= \lim_{\e\to0,\ \e>0}\, \dfrac{1}{z+i\e}\ .
\]
Then
\eq\ba{l}
\D_{AA}(1,2) = m\th_{12}^2 e^{i(\th_1\s\tb_2-\th_2\s\tb_1)\pa}
           \D_c(x_1-x_2)\ ,\es
\D_{A\AB}(1,2) = e^{i(\th_1\s\tb_2+\th_2\s\tb_1-\th_{12}\s\tb_{12})\pa}
           \D_c(x_1-x_2)\ ,\es
\D_{\AB\AB}(1,2) = m\tb_{12}^2 e^{i(\th_1\s\tb_2-\th_2\s\tb_1)\pa}
           \D_c(x_1-x_2)\ .
\ea\eqn{baby-prop-x}
Taking the Fourier transform with respect to $x_1-x_2$ yields the
momentum space propagators
\eq\ba{l}
\hat\D_{AA}(1,2) = m\th_{12}^2 e^{-(\th_1\s\tb_2-\th_2\s\tb_1)p}
           \hat\D_c(p)\ ,\es
\hat\D_{A\AB}(1,2) = e^{-(\th_1\s\tb_2+\th_2\s\tb_1-\th_{12}\s\tb_{12})p}
           \hat\D_c(p)\ ,\es
\hat\D_{\AB\AB}(1,2) = m\tb_{12}^2 e^{-(\th_1\s\tb_2-\th_2\s\tb_1)p}
           \hat\D_c(p)\ .
\ea\eqn{baby-prop-p}
\newpage
\section{Super Yang-Mills Theory}\label{sym}
This section contains a general description of the $N$=1 supersymmetric
gauge theories and of their gauge fixing procedure 
at the classical level. I follow~\cite{ps-book}, up to small changes in
the notation.

\subsection{Pure Super Yang-Mills Action}
The supermultiplet of gauge fields is given by the components of the
superfield (see \equ{theta-exp})
\eq\ba{l}
\f(x,\th,\tb) = C(x) + \th\chi(x) + \tb\bar\chi(x) + \half\th^2M(x)
+ \half\tb^2\bar M(x)  \es\phantom{\f(x,\th,\tb) =}
+ \th\s^\m\tb v_\m(x) + \half\tb^2\th\la(x)
+ \half\th^2\tb\bar\la(x) + \frac 1 4  \th^2\tb^2D(x)\ ,
\ea\eqn{gauge-s-f}
$\f$ as well as each of its components belong to the adjoint
representation of the gauge group $G$.
We use a matricial notation:
\eq
\vf=\vf^a\tau_a\ ,\quad \vf = \f,\ C,\chi,\cdots 
\eqn{matrix-not}
where the matrices $\tau_a$ form the basis of the Lie group $G$ in the defining
representation of $G$ -- e.g. the Pauli matrices in the case $G$ = SU(2)
-- normalized in such a way that
\eq
[\tau_a,\tau_b]=if_{abc}\tau_c\ ,\quad \tr\tau_a\tau_b=\d_{ab}\ .
\eqn{lie-alg}
The gauge transformations are implicitly defined by
\eq
e^{\f'} = e^{-i\bar\LA}e^\f e^{i\LA}\ ,\quad\mbox{with}\quad \DB_\ad\LA=0\ ,
\eqn{g-trf-phi}
where $\LA=\LA^a\tau_a$, 
which explicitly yields, for the infinitesimal transformations,
\eq\ba{l}
\d_{\rm gauge}\f = \dfrac{i}{2}L_\f(\LA+\bar\LA)+
   \dfrac{i}{2}\lp L_\f{\rm coth}(L_\f/2)\rp\lp\LA-\bar\LA\rp
\es
\phantom{\d_{\rm gauge}\f} = i(\LA-\bar\LA) +\dfrac{i}{2}[\f,\LA+\bar\LA]
  + \dfrac{i}{12}[\f,[\f,\LA-\bar\LA]] +O(\f^3)\ ,
\ea\eqn{inf-g-tr-phi}
with $L_\f X = [\f,X]$. 
 
\remark
Later we shall see that this transformation law is only a particular 
case of a general transformation law defined by
\eq
e^{\FF(\f')} =  e^{-i\bar\LA}e^{\FF(\f)}e^{i\LA}\ ,
\eqn{-gen-inf-g-tr-phi}
where $\FF(\f)$ is an arbitrary function of $\f$, only restricted by the
requirement to be in the adjoint representation like $\f$.
\kramer
The group G will be supposed to be a simple Lie group. Generalization to
a general compact Lie group is straightforward.
The pure super Yang-Mills (SYM) action reads~\cite{sym-orig} (the
conventions are those of~\cite{ps-book})
\eq\ba{l}
\S_{\rm SYM} = -\dfrac{1}{128g^2}\tr\is F^\a F_\a\ ,\\[4mm]
\quad \mbox{with}\quad F_\a = \DB^2\lp e^{-\f}D_\a e^\f\rp\ .
\ea\eqn{sym-action}

\subsection{The Wess-Zumino Gauge:}\label{jauge de wz}
Expanding in components the chiral superfield $\LA$ 
according to \equ{comp-chiral}:
\eq
\LA(x,\th,\tb) =
  e^{-i\th\s^\m\tb\pa_\m}\lp a(x) + \th\eta(x) + \th^2 f(x)\rp\ ,
\eqn{lambda}
one can write the gauge transformations for the components of the gauge
superfield $\f$ as
\eq\ba{l}
C'=C+i(a-\bar a)+ \cdots\ ,\quad
\chi'=\chi+i\eta+\cdots\ ,\quad M' = M +2if+\cdots\ ,\es
v'_\m=v_\m+\pa_m(a+\bar a)+\cdots\ ,\quad
\la'=\la+\sb^\m\pa_\m\eta+\cdots\ ,\es 
D'=D-i\pa^2 (a-\bar a)+\cdots\ . 
\ea\eqn{g-trf-components}
where the dots stand for the non-Abelian part of the transformations.
One can see that the transformations of the lower components 
$C$, $\chi$ and $M$ do not involve any derivative of the components of
$\LA$. It follows that one can solve algebraically 
for ${\rm Im}\, a$, $\eta$ and $f$ the equations
$C'=\chi'=M'=0$. Thus there always
exist a gauge transformation which allows to fix to zero these lower
components of $\f$. This defines the Wess-Zumino gauge~\cite{31}. 
In this gauge
only the higher components, i.e. the gauge field $v_\m$, the ``gaugino''
$\la$ and the $D$-field, are non-zero. From the components of $\LA$, 
only ${\rm Re}\,a$
remains free. It corresponds to the usual gauge transfornations:
\eq
A'_\m= e^{-i\om}(\pa_\m+i[A_\m,\om])e^{i\om}\ ,\quad
\la'= e^{-i\om}\la e^{i\om}\ ,\quad
D'= e^{-i\om}D e^{i\om}\ ,
\eqn{ord-g-trf}
where one has set
\eq
\om:={\rm Re}\,a   \ ,\quad A_\m:=\half v_\m\ .
\eqn{def-omega-a}
The SYM action \equ{sym-action} now reduces to the more familiar one
\eq
\S_{\rm SYM\ (WZ gauge)} = \dfrac{1}{g^2} 
\tr\xint\lp -\dfrac{1}{4}F^{\m\n}F_{\m\n}
+i\la\s^\m D_\m\bar\la
+\dfrac{1}{2}D^2 \rp\ ,
\eqn{sym-wz-action}
with
\[
F_{\m\n} = \pa_\m A_\n-\pa_\n A_\m + i[A_\m,A_\n]\ ,
\quad D_\m\cdot = \pa_\m\cdot+i[A_\m,\cdot] .
\]
Of course, the Wess-Zumino gauge is not preserved by the supersymmetry
transformations \equ{transf-fi}. However, the action
\equ{sym-wz-action} is still invariant under the following combination 
of infinitesimal supersymmetry and gauge transformations :
\eq\ba{l}
\d_\a A_\m = \dfrac{1}{4} (\s_\m\bar\la)_\a ,\es
\d_\a\la^\b= \d_\a^\b D + 2\s^{\m\n}F_{\m\n}\ ,\es
\d_\a\bar\la_\ad = 0\ ,\es
\d_\a D = -i(\s^\m D_\m\bar\la)_\a\ .
\ea\eqn{wzgauge-susy-trf}
These transformations are nonlinear, which is a source of
complications for the renormalization~\cite{ren-wz-gauge}. Moreover,
the supersymmetry algebra closes on the ``covariant translations'', instead
of the simple translations as in \equ{alg-delta-susy}: one has to
replace the derivative $\pa_\m$ in the translation operator by the
covariant derivative $D_\m$, when acting on $\la$ and $D$, 
and replace $\pa_\m A_\n$  by $F_{\m\n}$. The reader may
consult~\cite{white-etc} for recent progress in this direction.

\subsection{Gauge Fixing and BRS Invariance}
For the rest of these lectures, we shall choose a supersymmetric gauge
fixing, instead of the Wess-Zumino one described in the preceding
subsection. This gauge fixing will be a supersymmetric extension of the
Lorentz gauge $\pa^\m v_\m=0$. Observing that $\pa^\m v_\m$ is a
component of the chiral superfield
\eq
\DB^2D^2\f = e^{-i\th\s^\m\tb\pa_\m}\lp 
 4(D-\pa^2C -2i\pa v) - 8i\th\s\pa(\bar\la+i\pa\chi\s)
        -8\th^2\pa^2M \rp \ , 
\eqn{long-s-field}
we shall implement the condition $\DB^2D^2\f=0$, with the help of a
Lagrange multiplier chiral superfield $B=B^a \tau_a$. 
We thus add to the action the piece 
\[
\dfrac{1}{8}\tr\is B\DB^2D^2\f + \mbox{ c.c.}
= \dfrac{1}{8}\tr\iv (B D^2\f + \BBAR \DB^2\f)\ .
\]
Since the gauge group is non-Abelian one has still to add Faddeev-Popov 
ghost fields. The gauge condition and the gauge parameter
$\LA$ being chiral, these ghost fields will be chiral as well. 
We note
that $c_-=c_-^a \tau_a$ and $c_+=c_+^a \tau_a$. 
They are the antighost and the ghost, respectively. Their components 
$c_\pm$ and $\cb_\pm$ are {\it anticommuting} or Grassmann chiral
superfields. 

Before introducing them in the action, let us define
the BRS transformations, under which the total action will have to be
invariant:
\eq\ba{l}
s\f = \dfrac{1}{2}L_\f(c_+ +\cb_+)+
   \dfrac{1}{2}\lp L_\f{\rm coth}(L_\f/2)\rp\lp c_+ -\cb_+\rp\es
\phantom{s\f} = c_+ - \cb_+ + \half[\f,c_+ + \cb_+] + \cdots\ ,\es
s c_+ = -c_+^2\ ,\quad s \cb_+ = -\cb_+^2\ ,
  \quad\lp\ sc_+^a=-\dfrac{i}{2} f_{abc}c_+^b c_+^c\ \rp \es
sc_- = B\ ,\quad s\cb_-=\BBAR\ ,\es
sB=0\ ,\quad s\BBAR = 0\ .
\ea\eqn{brs-transf}
One checks that the BRS operator $s$ is an {\it antiderivation}
which is {\it nilpotent}:
\eq
s^2=0\ .
\eqn{nilpotency}
One sees that the BRS transformation of the gauge superfield $\f$ is
just the gauge transformation \equ{inf-g-tr-phi} -- up to a factor $i$.
The gauge invariant action \equ{sym-action} thus is already BRS
invariant. The gauge fixing piece of the action will be defined as
\eq\ba{l}
\S_{\rm gf} =  \dfrac{1}{8} s\,\tr\iv
      \lp c_- D^2\f + \cb_- \DB^2\f\rp \es
\phantom{\S_{\rm gf}}
 = \dfrac{1}{8}\tr\iv \lp B D^2\f + \BBAR \DB^2\f
 - c_- D^2s\f -\cb_- \DB^2s\f \rp\ .
\ea\eqn{gauge-fix}
Its BRS invariance follows from the nilpotency of $s$. The last term,
which involves the ghosts and the antighosts, is the supersymmetric
extension of the usual Faddeev-Popov action. 
\remark
One may add to the gauge fixing action a supplementary term
\eq
\S_{(\a)} = 2\a \,\tr\iv B\BBAR\ ,
\eqn{action-alpha}
where $\a$ is a  dimensionless {\it gauge parameter}. 
One can show~\cite{ps-book} that the physical content of 
the theory does not depend on it.
This makes of the $B$ field  an auxiliary field which can be eliminated
by using its equation of motion
\eq
\DB^2 \BBAR = -\dfrac{1}{16\a}\DB^2 D^2\f \ ,
\eqn{alpha-b-eq-motion}
thus yielding, for the $B$-dependent terms of the action the expression
\eq
-\dfrac{1}{256\a}\tr\iv \f \lp D^2\DB^2+\DB^2D^2\rp\f\ ,
\eqn{stuck-gauge}
which is the supersymmetrization of the Stueckelberg gauge fixing
\[
-\dfrac{1}{2\a}\tr\xint \lp\pa^\m v_\m\rp^2\ .
\]
But, using the fact that the physical quantities are independent from 
$\a$~\cite{gauge-ind}, we shall keep $\a=0$ 
through the rest of these lectures. This
corresponds to a supersymmetrization of the Landau gauge fixing
\[
\tr\xint \lp{\rm Re}\,F_B\rp \pa^\m v_\m ,
\]
where $F_B$ is the $\th^2$-component of the chiral superfield $B$.
\kramer

\subsection{Matter Fields}
Having written all the pieces building the classical gauge fixed 
action of the pure super Yang-Mills action, let us introduce matter. 
The latter is described by a set of chiral superfields
\eq
A^i(x,\th,\tb) = 
  e^{-i\th\s^\m\tb\pa_\m}\lp A^i(x) + \th\p^i(x) + \th^2 F^i(x)\rp\ ,
\eqn{matter-fields}
which belong to some representation $R$ of the gauge group. 
Their BRS transformations -- identical to their infinitesimal gauge
transformations up to a factor $i$ -- read
\eq
sA^i = -c_+^a T_a{}^i{}_j A^j \equiv -(c_+A)^i  \ ,\quad 
  s\AB_i =  \AB_j T_a{}^j{}_i \cb^a_+ \equiv (\AB \cb_+)_i \ ,
\eqn{brs-matter}
where the hermitean matrices $T_a$ are the generators of the gauge
group in the representation $R$. 

The BRS-invariant action for the matter fields reads
\eq
\S_{\rm matter} = \dfrac{1}{16} \iv \AB e^{\f^a T_a}A + \is W(A)
  + \isb \bar W(\AB)\ ,
\eqn{matter-action}
with the superpotential $W$ given by
\eq
W(A) = \dfrac{1}{8}m_{(ij)}A^iA^j + \la_{(ijk)}A^iA^jA^k\ ,
\eqn{matter-s-pot}  
the mass matrix $m_{ij}$ and the Yukawa coupling constants $\la_{ijk}$
being invariant symmetric tensors in the representation $R$.

\subsection{$R$ Invariance}\label{invariance R}
 It is easy to check that, in the massless case ($m_{ij}=0$), 
the classical action given by
\equ{sym-action}, \equ{gauge-fix} and \equ{matter-action} is invariant
under the $R$-transformations generically defined by \equ{r-transf}, 
the $R$-weights $n$ of the various superfields of the present theory 
being given in Table~\ref{table1}.

This symmetry will play a very important role in the sequel.
\begin{table}[hbt]
\centering
\begin{tabular}{|c||c|c|c|c|c|c|c|c|c|c|}
\hline
    &$\th^\a$ &$D_\a$ &$\f$ &$A$ &$c_+$ &$c_-$ &$B$ &$\f^*$ &$A^*$ 
     &$c_+^*$ \\
\hline\hline
$d$ &$-{1\over2}$ &${1\over2}$ &0 &1 &0 &1 &1 &2 &2 &3 \\
\hline
$n$ &$-1$ &1 &0 &$-{2\over3}$ &0 &$-2$ &$-2$ &0 &$-{4\over3}$ &-2  \\
\hline
$\F\Pi$&0 &0 &0 &0 &1 &$-1$ &0 &$-1$ &$-1$ &$-2$ \\
\hline
\end{tabular}
\caption[t1]{Dimensions  $d$, R-weights $n$ and  ghost numbers $\F\Pi$.}
\label{table1}
\end{table}

\subsection{Slavnov-Taylor identity}
In order to express the BRS invariance of the theory through 
a Ward identity, we have to take care of the nonlinearity of 
the BRS transformations. We couple the latters with external superfields
$\f^*$, $A^{*i}$ and $c_+^*$ by adding to the action the piece 
\eq
\S_{\rm ext} =  \iv\tr \f^*s\f + \lc \is\lp A^{*i}sA_i+\tr
c_+^*sc_+\rp  + {\rm c.c.} \rc
 \equiv \dint\dsum{\vf}{} \vf*s\vf  \ .
\eqn{ext-action}
The BRS invariance of the total action
\eq
\S := \S_{\rm SYM}+ \S_{\rm matter} + \S_{\rm gf} + \S_{\rm ext}
\eqn{tot-action}
is now expressed by the Slavnov-Taylor identity
\eq
\SS(\S)= 0\ ,
\eqn{g-class-slavnov}
with 
\eq\ba{l}
\SS(\g) = \tr \dint d V {\dfrac{\d \g}{\d \f^*}} {\dfrac{\d \g}{\d \f}}
+ \lp \dint d S \left\{ {\dfrac{\d \g}{\d A^{*i}}} {\dfrac{\d \g}{\d A_i}}
+ \tr {\dfrac{\d \g}{\d c_+^*}} {\dfrac{\d \g}{\d c_+}} + \tr B
{\dfrac{\d \g}{\d c_-}} \right\} + {\rm c.c.} \rp \es
\phantom{\SS(\g)}\equiv \dint\dsum{\vf}{}\dfud{\g}{\vf^*}\dfud{\g}{\vf}
   + B\dfud{\g}{c_-} \ .
\ea\eqn{8.7}
The gauge fixing is defined in a functional way by the 
condition {}({}supersymmetric Landau gauge)
\eq
{\dfrac{\d \S}{\d B}} = {\dfrac{1}{8}} \bar{D}^2 D^2 \f \ ,
\eqn{8.8}
and its complex conjugate.

Differentiating the Slavnov-Taylor 
identity with respect to $B$ or $\bar{B}$ and
using \equ{8.8} yield the ghost equations
\eq
\GG_+ \S = 0 \ , \quad \bar{\GG}_+\S = 0 \ ,
\eqn{8.9}
with
\eq
\GG_+ = {\dfrac{\d}{\d c_-}} + {\dfrac{1}{8}} \bar{D}^2 D^2
{\dfrac{\d}{\d \f^*}} \ .
\eqn{8.10}
They imply that the theory depends on $c_-$, $\bar{c}_-$ and
$\f^*$ only through the combination
\eq
\hat{\f^*} = \f^* - {\dfrac{1}{8}} (D^2 c_- + \bar{D}^2 \bar{c}_-)\ .
\eqn{8.11}
 Hence
\eq\ba{l}
\S = \S(\f,A,c_+,B,\hat{\f^*},A^*,c_+^*)\es
\phantom{\S} = \S_{\rm SYM}(\f) + \S_{\rm matter}(\f,A)
 + \S_{\rm ext}(\f,A,c_+,\hat{\f^*},A^*,c_+^*)  \es
\phantom{\S=}+ \dfrac{1}{8} \tr\dint dV (BD^2\f+\bar{B}\bar{D}^2\f) \ ,
\ea\eqn{8.111}
\subsection{Rigid Invariance}
The total action is invariant under the rigid transformations 
\eq\ba{l}
\d_{\rm rig} \vf = i [\om,\vf]\ ,\quad \vf = 
   \f,\,c_\pm,\,B,\,\f^*,\,c_+^*\ ,\es
\d_{\rm rig} A^i = i\om^a T_a{}^i{}_j A^j \ , \quad
\d_{\rm rig} A^*_i = -i\om^a A^*_j T_a{}^j{}_i \ ,
\ea\eqn{rig-transf}
which correspond to gauge
transformations with constant parameters $\om^a$. 

Rigid invariance does not necessarily hold in general. It holds
here because the gauge fixing condition respects it. This would not be
the case with a more general gauge fixing condition, such as
a 't Hooft-like gauge, for example. 

\subsection{Ward Identities and Algebra}
Beyond BRS invariance, the theory posesses invariances under
supersymmetry, translations, $R$-transformations
and rigid transformations. The four latter
symmetries being linear are expressed by the Ward identities
\eq
W_X\S:= -i \dsum{\vf}{} \dint \d_X\vf\dfud{}{\vf}\S=0\ ,
\quad X= Q_\a,\, P_\m,\, R, \mbox{ and rigid transf.}   \ ,
\eqn{sym-w-i}
where $\d^P_\m\vf=\pa_\m\vf$, and
 $\d^Q_\a$, $\d^R$, $\d_{\rm rig}$ 
are defined by  \equ{delta-susy}, \equ{r-transf},
 \equ{rig-transf}, respectively. 

The Ward identity operators together with the Slavnov-Taylor operator
 and the gauge fixing and ghost equation operators,
obey the algebra (null (anti-)commutators are not written) 
\eq\ba{l}
\lc W^Q_\a, W^{\bar Q}_\ad \rc = 2\s^\m_{\a\ad} W^P_\m\ ,\quad
  \lc W^Q_\a,W^R \rc = -W^Q_\a\ ,\quad \lc W^Q_\ad,W^R \rc 
   = +W^Q_\ad\ ,\es
W^Q_\a \SS(\g) - \SS_\g W^Q_\a\g = 0\ ,\quad
   W^R \SS(\g) - \SS_\g W^R\g = 0 \ ,\quad
   W_{\rm rig}\SS(\g) - \SS_\g W_{\rm rig}\g = 0\ ,\quad\forall\,\g\ ,\es
\fud{}{B}\SS(\g) - \SS_\g\lp\fud{}{B}\g-\frac{1}{8} \DB^2 D^2 \f\rp
  = \GG_+\g\ , \quad\forall\,\g\ ,\es
\GG_+\SS(\g) + \SS_\g\GG_+\g = 0\ , \quad\forall\,\g\ ,\es
\SS_\g\SS(\g)=0\ ,\quad\forall\,\g\ ,\es
{\SS_\g}^2 = 0\quad\mbox{if}\quad\SS(\g)=0\ ,
\ea\eqn{alg-ward-slavnov}
$\g$ denoting a functional of the superfields and $\SS_\g$ the
``linearized'' Slavnov-Taylor operator at the ``point'' $\g$:
\eq\ba{rl}
\SS_\g =& \tr \dint d V \lp 
{\dfrac{\d \g}{\d \f^*}} {\dfrac{\d}{\d \f}} +{\dfrac{\d \g}{\d
\f}} {\dfrac{\d}{\d\f^*}} \rp    \es
&+\lp \dint d S \lp {\frac{\d \g}{\d
A^*}} {\dfrac{\d}{\d A}} + {\dfrac{\d \g}{\d A}} {\dfrac{\d}{\d A^*}}
+ \tr {\dfrac{\d \g}{\d c_+^*}} {\dfrac{\d }{\d c_+}} + \tr
{\dfrac{\d \g}{\d c_+}} {\dfrac{\d}{\d c_+^*}}
  + \tr B\dfud{}{c_-}\rp + {\rm c.c.}\rp  \es
\ea \eqn{lin-slavnov}
Note that, since the classical action $\S$ obeys the Slavnov-Taylor
identity, $\SS_\S$ is nilpotent:
\eq
\SS_\S{}^2 = 0\ .
\eqn{nilp-s-sigma}

\subsection{General Classical Action}\label{solution generale}
The general solution of the classical problem, i.e. of solving the
Slavnov-Taylor identity for the classical action, taking into account 
the gauge condition \equ{8.8} and the Ward identities \equ{sym-w-i} for
supersymmetry, $R$-invariance and rigid invariance, is given by
\eq\ba{l}
\S(\f, A, c_+,B, \hat{\f^*}, A^*, c_+^*) = 
\S_{\rm SYM} (\f ') + \S_{\rm matter} (\f ' , A') +
\S_{\rm ext} (\f', A', c_+', \hat{\f^*}', {A^*}', {c_+^*}')  \es
\phantom{\S(\f, A, c_+,B, \hat{\f^*}, A^*, c_+^*) = }
+  {\dfrac{1}{8}} \tr \dint d V ( B D^2 \f +  \bar{B} \bar{D}^2 \f) \ ,
\ea\eqn{8.16}
with
\eq\ba{ll}
\f'  =  {\cal F} (\f)\ , 
&\quad {\hat{\f^*}}'(z)  = \left.\dfud{}{\f'(z)}\dint dV(z') 
  \hat{\f^*}(z') \FF^{-1}(\f')(z') \right|_{\f'=\FF(\f)} \ ,  \es
A'  =  z_1 A  \ , &\quad {A^*}'= \dfrac{1}{z_1} A^* \ ,   \es
c_+'  =  z_2 c_+ \ ,& \quad {c_+^*}' = \dfrac{1}{z_2} c_+^*  \ .
\ea\eqn{8.18}
Eqs. \equ{8.18} represent field renormalizations. Due to the
dimensionlessness of $\f$ its renormalization is non-linear: the
function $\FF (\f )$ is an arbitrary formal power series in $\f$:
\eq
\FF_a(\f) = \sum a_k \ t^{(k)}_{a a_1 \cdots a_n} \f^{a_1} \cdots
\f^{a_n}
\eqn{8.19}
Due to the rigid invariance \equ{rig-transf}, the numbers
$t_{a a_1 \cdots a_n}^{(k)}$ are  components of invariant
tensors of the group. (In the same way the masses $m_{ij}$ and
the couplings $\la_{ijk}$ in \equ{matter-action} 
are invariant tensors in the
matter field representation). We shall restrict ourselves in the
following to the massless case $m_{ij} = 0$.

One can check that the dependence on the renormalization
parameters $a_k$ is non-physical. This is expressed by the fact
that the derivative of the action with respect to each $a_k$ is
a BRS-variation:
\eq
\dpad{}{a_k} \S = \SS_{\S} \D_k  \ ,
\eqn{8.20}
where $\D_k$ is some local functional. This means that the $a_k$ are gauge
parameters~\cite{gauge-ind}.
\remark
{\small Let us open a ``parenthesis'': 
The gauge condition and the ghost equation lead to the decomposition
\eq
\S= \hat{\S} (\f,\ A,\ c_+,\ \hat{\f^*},\ A^*, c_+^*) + 
 {\dfrac{1}{8}} \tr \dint d V ( B D^2 \f +  \bar{B} \bar{D}^2 \f) \ .
\eqn{8.12}
The Slavnov-Taylor identity then reads
\eq
\SS(\S) = \half \BB_{\hat{\S}} \hat{\S} = 0\ ,
\eqn{8.13}
with
\eq\ba{rl}
\BB_\g =& \tr \dint d V \lp 
{\dfrac{\d \g}{\d \f^*}} {\dfrac{\d}{\d \f}} +{\dfrac{\d \g}{\d
\f}} {\dfrac{\d}{\d\f^*}} \rp    \es
&+\lp \dint d S \lp {\frac{\d \g}{\d
A^*}} {\dfrac{\d}{\d A}} + {\dfrac{\d \g}{\d A}} {\dfrac{\d}{\d A^*}}
+ \tr {\dfrac{\d \g}{\d c_+^*}} {\dfrac{\d }{\d c_+}} + \tr
{\dfrac{\d \g}{\d c_+}} {\dfrac{\d}{\d c_+^*}}\rp + {\rm c.c.}\rp  
\ea \eqn{8.14}
obeying the identities
\eq\ba{l}
{\cal B}_\g {\cal B}_\g \g  =  0\ , \qquad \forall\, \g \ ,    \es
{\cal B}_\g {\cal B}_\g     =  0 
               \qquad {\rm if} \qquad {\cal B}_\g \g = 0 \ .
\ea\eqn{8.15}
}\kramer

\subsection{Soft Breakings of Supersymmetry}
If supersymmetry has some relevance, it must be broken at ``low'' energy
(typically below $\approx 1$Tev). A spontaneous breakdown is conceivable at
the level of supergravity, i.e. of local supersymmetry~\cite{sugrav}, or at
the level of superstring theory~\cite{s-string}.

But, in the low energy domain, where gravitational 
interaction appears to be
negligible, i.e. in the domain of {\it rigid} supersymmetry, the
breakdown arises in the form of an explicit breakdown by nonsupersymmetric
mass terms. Such a breakdown is soft{\footnote{``soft'' is taken here in
the sense of power-counting~\cite{symanzik}. This definition is more general
that the one of \cite{girardello-g}, which only keeps 
breakings which do not give rise
to UV divergences more severe than logarithmic.}, which means 
that it does not affect
the behaviour of the theory in the high energy domain, where supersymmetry
thus remains valid.

Such soft breakings are conveniently described in the 
Symanzik formalism~\cite{symanzik}, 
which consists in adding to the action couplings with ``shifted''
fields (here: shifted superfields) in order to keep record of the
transformation properties of the breaking terms, in such a way that the Ward
identities still hold. Let us see an example of breaking by
a gauge invariant gluino mass term $\tr(\la^\a\la_\a)$. 
One observes that
\eq
\tr \la^\a\la_\a= \tr F^\a F_\a |_{\th = 0}\ ,
\eqn{gluino-mass}
i.e., this term is equal to the first component of the chiral superfield 
$\tr F^\a F_\a$
(where $F_\a$ is given by (\ref{sym-action})).

\noindent One then introduces the shifted chiral superfield
\eq
E'=E+m\th^2
\eqn{shift}
where $E$ is a gauge invariant  classical 
chiral superfield, and $m$ is a parameter with the dimension of a mass.
The term 
\eq
\dint dS E' \tr F^\a F_\a
\eqn{e-prime}
just gives the gluino mass term \equ{gluino-mass}
at $E=0$. Moreover the action  containing this term still obeys
the supersymmetry Ward identity
\eq
W^Q_\a\S=0\ ,
\eqn{e-prime-wi}
where 
\eq
W^Q_\a=W^{Q\,\rm (old)}_\a + \lc\is \d^Q_\a E' {\dfrac{\d }{\d E'}} 
+ c.c. \rc\ ,
\eqn{e-prime-wi-op}
where $\d^Q_\a E'$ resuts from the application of the supersymmetry 
differential generator given by eq. (\ref{def-delta-susy}) 
of Appendix~\ref{notations} to $E'$:
\eq
\d^Q_\a E'= \d^Q_\a E + 2m \th_\a\ .
\eqn{susy-e'}
One easily sees that, at $E=0$, the Ward identity (\ref{e-prime-wi}) reads
\eq
W^{Q\,\rm (old)}_\a\S=-\xint m\tr (\la_\a D + \dots)\ .
\eqn{susy-wi-shift}
The right-hand side is, as one expects, the variation of the spacetime
integral of the gluino masss term 
\equ{gluino-mass}, computed using the transformation rules 
\equ{transf-fi} of Appendix~\ref{notations}. 
The advantage of the supersymmetric formalism is that the algebra 
\equ{wi-op-alg} is preserved by the Ward operator 
involving such shifted superfields.

\newcommand{\dc}{{\D^{\rm c}}}
\newcommand{\tti}{{\tilde\th}} 
\newcommand{\thpp}{{\th_{\cdot\cdot}}} \newcommand{\tbpp}{{\tb_{\cdot\cdot}}} 
\newcommand{\ttipp}{{\tti_{\cdot\cdot}}} 
\newpage
\section{Superspace Feynman Graphs}\label{superdiagrammes}
A short account of the supergraph formalism and of its consequences
will be given. A more detailed account is given in~\cite{ps-book}. 
See~\cite{gates-book} for a somewhat alternative presentation.
\subsection{The Free Propagators}
The free propagators are the Green functions of the theory defined by
the quadratic part of the classical action, the fields 
$\vf_i(x)$ being coupled to external sources
$J^i(x)$, ($i=1,\cdots,n$). Let us first illustrate the
procedure for obtaining them in the case the $\vf^i$'s are 
scalar fields. The field equations corresponding to the action
\eq
\S_{\rm free}(\vf) = \dint dx\lp \half \vf_i K^{ij}\vf_j + J^i\vf_i \rp\ ,
\eqn{fey-scalar-action}
where $K^{ij}$ is a matrix of partial derivative operators, read
\eq
K^{ij}\vf_j + J^i = 0\ .
\eqn{fey-scalar-eq}
The Green functions are defined as the solutions of the equations
\eq
K^{ij}\dc_{jk} (x) = i\d^i_k\d(x-y)\ ,
\eqn{fey-prop-eq}
with the Feynman-Stueckelberg causal prescription as a boundary condition.
Then, the solution of the field equations \equ{fey-scalar-eq} will read
\eq\ba{l}
\vf_i(x) = i\dint dx \dc_{ij} (x-y) J^j(y) \ ,\es
\mbox{or, formally:}\quad\quad
  \vf = i\dc*J\ ,\quad\mbox{with}\quad \dc = \dfrac{i}{K}  \ .
\ea\eqn{fey-sol-scalar}
E.g., in the case of one scalar field, $K=-\pa^2-m^2$,
\eq\ba{l}
\dc = \dfrac{-i}{\pa^2+m^2} = \dfrac{1}{(2\pi)^4}\dint dp\,
 e^{ip(x-y)}{\tilde\D}^{\rm c}(p)\ , \es
\mbox{with}\quad \dc(p) = \dfrac{i}{p^2-m^2+i0}
  =: \lim_{\e\to+0}\dfrac{i}{p^2-m^2+i\e}\ .
\ea\eqn{fey-scalar-prop}

Let us go to the super Yang-Mills theory\footnote{We consider the
massless theory, leaving the massive case as an exercise to the
reader.} beginning with the matter
fields. The free action being diagonal we don't write the summation 
indices:
\eq
\S_{\rm free}(A) = \dfrac{1}{16}\iv \AB A + \is AJ+\isb \AB\JB\ .
\eqn{fey-matter-action}
The corresponding field equations read
\[
\dfrac{1}{16}\DB^2\AB = -J\ ,\quad \dfrac{1}{16}D^2A = -\JB\ .
\]
Applying the operator $\DB^2$ to the second equation, and using the
commutation relations \equ{com-cov-der} of Appendix~\ref{notations}, we
find the equation 
\[
\pa^2A = \DB^2 \JB\ ,
\]
whose solution reads\footnote{The superspace 
integration measure used in the
convolution product $*$ is the one appropriate to the type of the
superfields involved in each case.}
\[
A = i\dc_{AA}*J + i\dc_{A\AB}*\JB = \dfrac{1}{\pa^2} \DB^2\JB\ .
\]
This yields
\eq\ba{l}
\dc_{AA}(x_1,\th_1,\tb_1;x_2,\th_2,\tb_2)\equiv\dc_{AA}(1,2)= 0\ ,\es
\dc_{A\AB}(x_1,\th_1,\tb_1;x_2,\th_2,\tb_2)\equiv\dc_{A\AB}(1,2)
 = \dfrac{-i}{\pa^2}\DB^2\d_\SB(1,2) \es
\quad = \dfrac{1}{(2\pi)^4}\pint{\tilde\D}^{\rm c}_{A\AB}(p;1,2)\ ,
  \quad\mbox{with}\quad {\tilde\D}^{\rm c}_{A\AB}(p;1,2) = 
 e^{-(\th_1\s\tb_2-\th_2\s\tb_1-\th_{12}\s\tb_{12})p} \dfrac{i}{p^2+i0}\ ,
\ea\eqn{fey-mat-prop}
where we have used the expressions \equ{delta-funct}, \equ{dd-delta} of
Appendix~\ref{notations} for
the superspace Dirac distributions and their derivatives.

The free action for the ghost fields reads
\[\ba{l}
\S_{\rm free}(c_+,c_-) = -\dfrac{1}{8}\iv c_- D^2 c_+  
 + \is \lp J_{c_+}c_+ + J_{c_-}c_-\rp \quad+\ \mbox{c.c.}\es
\phantom{\S_{\rm free}(c_+,c_)} = \is\lp 2 c_- D^2 c_+
 +  J_{c_+}c_+ + J_{c_-}c_-\rp \quad+\ \mbox{c.c.}
\ea\]
From the field equations
\[
2\pa^2c_+ = J_{c_-} \ ,\quad -2\pa^2c_- = J_{c_+} \ ,
\]
we immediately deduce the ghost propagators (directly written in
momentum space):
\eq\ba{l}
{\tilde\D}^{\rm c}_{c_+c_-}(p;1,2)= 
  e^{-(\th_1\s\tb_2-\th_2\s\tb_1)p} (\th_{12})^2 \dfrac{i}{8(p^2+i0)}\es
{\tilde\D}^{\rm c}_{c_+\cb_-}(p;1,2)= 0 \ .
\ea\eqn{fey-ghost-prop}
In the sector of the gauge and Lagrange-multiplier fields the free
action is again diagonal in the Yang-Mills indices:
\[\ba{l}
\S_{\rm free}(\f,B) = \iv\lp -\dfrac{1}{128g^2}D\f\DB^2D\f
  +\dfrac{1}{8}(BD^2\f + \BBAR\DB\f) \rp \es
\phantom{\S_{\rm free}(\f,B) =} + \iv J_\f\f+ \is J_B B 
  + \isb J_\BBAR \BBAR\ .
\ea\]
It yields the field equations
\eq\ba{l}
\dfrac{1}{64g^2} D\DB^2D\f + \dfrac{1}{8} D^2B  
 + \dfrac{1}{8} \DB^2\BBAR = -J_\f \ , \es
\dfrac{1}{8} \DB^2D^2\f = -J_B\ ,\es
\dfrac{1}{8} D^2\DB^2\f = -J_\BBAR\ .
\ea\eqn{fey-b-phi-eqs}
Applying $\DB^2$ on the first equation we obtain
\[
\dfrac{1}{8} \DB^2D^2 B = -\DB^2 J_\vf\ ,
\]
which solves into
\[
B = \half\dfrac{1}{\pa^2}\DB^2 J_\f\ .
\]
From the latter equation one reads out the propagators (in momentum
space): 
\eq\ba{l}
{\tilde\D}^{\rm c}_{BB}={\tilde\D}^{\rm c}_{B\BBAR} = 0\ ,\es
{\tilde\D}^{\rm c}_{B\f}(p;1,2) = e^{-(\th_1\s\tb_2-\th_2\s\tb_1)p}
   (\th_{12})^2\dfrac{i}{8(p^2+i0)}\ .
\ea\eqn{fex-b-phi-prop}
In order to find the remaining propagators, we act on the first of the
equations \equ{fey-b-phi-eqs} with the ``transverse'' projector 
$P^{\rm T}$ \equ{projectors}, and sum up the
the second and the third of these equations.
Combining both equations thus obtained, making use of the
completeness property $P^{\rm T}+P^{\rm L}=1$ (see \equ{projectors}),
we obtain
\[
\dfrac{1}{8g^2}\pa^2\f =  -P^{\rm T}J_\f 
  +\dfrac{1}{16g^2}(J_B+J_\BBAR) \ ,
\]
which yields  
\eq
\dc_{\f\f}(1,2) = \dfrac{8ig^2}{\pa^2} P^{\rm T} \d_V(1,2)\ .
\eqn{fey-phi-prop-x}
\remark
Note that this yields, in momentum space, a double pole at $p^2=0$:
\eq
\dfrac{1}{(p^2+i0)^2}\ ,
\eqn{dipole}
which constitutes an infrared singularity.
We shall comment on this point at the beginning of
Section~\ref{renormalisation}.
\kramer
\subsection{Feynman rules and Power-Counting}
The contribution of an $L$-loop 
 superspace Feynman graph $\g$ consists in a product of superpropagators
or of covariant derivatives thereof. It has the form of an $L$-loop integral
\eq
J_\g(p,\tti) = \dint d^{4L}k\;I_\g(p,k,\tti)\ ,
\eqn{fey-integr}
where $\tti$ = $\th$ or $\tb$, and where
the $p$'s and the $k$'s denote the internal and external momenta,
 respectively. 
The precise structure of the integrand $I_\g$ follows from the following
momentum space Feynman rules (for the 1PI amputated diagrams):
\begin{enumerate}
\item[1.] For each internal line, write the corresponding
superpropagator, with appropriate derivatives if the vertices coupled by
the line involve superfield derivatives.
\item[2.] For each external (amputated) leg, write a superspace Dirac
distribution: $\d_V$, $\d_S$ or $\d_\SB$ (see \equ{delta-funct})
according to the nature of the
field (real, chiral or antichiral) associated to the leg.
\item[3.] At each vertex integrate over its  $\tti$ variables, with the
integration measure which corresponds to the nature of the vertex. 
\end{enumerate}
As we have seen in the preceding subsection,
the propagators as well as their covariant derivatives
have the structure
\eq
\dc(p;1,2) = e^{-(\th_1\s\tb_2 - \th_2\s\tb_1)p}f(p,\tti_{12})\ ,
\eqn{fey-gen-prop}
where $\tti_{ij}=\tti_i-\tti_j$. The superspace Dirac distributions have
the same structure.
It follows that the integrand $I_\g$ has the general form
\eq\ba{l}
I_\g(p,k,\tti) = e^{E(p,\tti)} \bar I_\g(p,k,\ttipp)\ ,\es
\mbox{with}\quad E(p,\tti) = 
   -\dsum{i=1}{n-1}(\th_i\s\tb_n-\th_n\s\tb_i)p_i\ ,
\ea\eqn{fey-gen-integrand}
where $\bar I_\g$ only depends on the differences $\ttipp$.
\remark
This corresponds to the general structure of a superfield Green function
following from the supersymmetry Ward identities:
\[
\tilde G(p,\tti) = e^{E(p,\tti)} \bar G(p,\ttipp) \ .
\]
\kramer\noindent
Expanding \equ{fey-gen-integrand} in powers of the $\tti$'s, we obtain
\eq
I_\g(p,k,\tti) = e^{E(p,\tti)} 
 \dsum{\om=0}{\OM}(\ttipp)^\om I_\om(p,k)\ ,
\eqn{exp-integrand}
where $(\ttipp)^\om$ stands for a generic monomial in the variables 
$\tti_{ij}$
of degree $\om$ -- a subsummation over all 
the independent monomials with the same degree being implicitly understood.
The maximum degree is given by
\eq
\OM = \lac \ba{ll}4N_V+2N_S+2N_\SB -4\quad 
                         &\mbox{in the generic case}\ ,\es
                  2N_S-2 &\mbox{if }N_V=N_\SB=0\ ,        \ea\right.
\eqn{omega-max}
where  $\ N_V$, $N_S$, resp. $N_\SB$  are the numbers of 
vector, chiral, resp. antichiral external legs of the 1PI graph
under consideration.
A simple argument based on dimensional analysis shows that 
the degrees of divergence $d_\om$ of 
the integrands  $I_\om$ are related to each
other by the formula\footnote{Due to the presence of possible mass terms,
such a powercounting based on dimensional analysis yields only 
upperbounds.} 
\eq
d_\om \le d_0+ \half\om\ .
\eqn{d-omega}
A detailed analysis\footnote{$d_0$ in fact is the degree of divergence
which one would obtain through usual power-counting for the component 
diagram whose external legs correspond to the highest $\th$-components
of the superfields coresponding to the legs of the superdiagram.}
leads for $d_0$ to the upperbound
\eq
d_0 \le 4 - \dsum{V}{}(d_V+2) - \dsum{S}{}(d_S+1) 
  - \dsum{\SB}{}(d_\SB+1) \ ,
\eqn{d-zero}
where $d_V$, $d_S$, resp. $d_\SB$ are the dimensions 
(see Table~\ref{table1} of in Section~\ref{sym}) of the superfields
corresponding to the vector, chiral, resp. antichiral external legs of
the diagram. 
The maximum degree of divergence of a supergraph is given by 
(use \equ{omega-max}):
\eq
d_\OM \le \lac\ba{ll} 2 - \dsum{V}{}d_V - \dsum{S}{}d_S - \dsum{\SB}{}d_\SB
      \quad &\mbox{in the generic case}\ ,\es
      3 - \dsum{S}{}d_S &\mbox{if }N_V=N_\SB=0\ . \ea \right.
\eqn{d-Omega}

\subsection{Nonrenormalization Theorem for the Superpotential}
Applying this result to the super Yang-Mills theory 
described in the preceding section, we find that the potentially
divergent diagrams (we don't consider here the diagrams with ghost
external legs) are those contributing to the following vertex 
functions~\footnote{A vertex functions is the sum of 
the contributions of the one-particle irreducible graphs 
only to a given Green functions, amputated from its external legs
(c.f. Appendix~\ref{fonctionelles}).}:
\eq\ba{ll}
\G_{\f_1\cdots\f_{N_V}}\ ,\quad\forall\, N_V\ :\quad& d_\OM \le 2\ ,\es
\G_{\f_1\cdots\f_{N_V}A\bar A}\ ,\quad\forall\, N_V\ :\quad
                                         & d_\OM \le 0\ ,\es
\G_{A_1A_2}\ : & d_\OM \le 1\ ,\es
\G_{A_1A_2A_3}\ : & d_\OM \le 0\ .
\ea\eqn{div-diagr}

But it turns out that the actual degrees of divergence are lower. In particular,
those for the purely chiral vertex functions $\G_{AA}$ and $\G_{AAA}$
are negative: the corresponding diagrams are convergent. This is the content 
of {\it the nonrenormalization theorem for the chiral vertices}. 
This theorem follows 
from the vanishing of the radiative (i.e. loop graph) corrections to the
purely chiral vertex functions at zero external momenta:
\eq
\left.\G^{\rm (rad.\ corr.)}_{A_1\cdots A_N}(p)\right\vert_{p=0} = 0 \ .
\eqn{p=0-chir-vert}
The latter result indeed implies that these radiative corrections
 must have external momentum factors, 
of degree 2 at least due to Lorentz invariance: 
hence their effective degree of divergence is lowered by 2 at least, 
which makes it negative in view of \equ{div-diagr}.
Before proving \equ{p=0-chir-vert}, let us note that 
this means that the effective superpotential 
defined (in momentum space) as
\eq
W_{\rm eff}(A) = \left.\G(\vf)\right|_{\ {\rm at\ zero\ momenta}, \ 
               \vf=0\ \forall\vf\ {\rm except\ }\vf=A } \ ,
\eqn{eff-superpot}
does not get any quantum correction:
\eq
W_{\rm eff}(A) = W_{\rm class}(A)\ ,
\eqn{nonren-superpot}
where $W_{\rm class}(A)$ is the classical superpotential, given by 
\equ{baby-s-pot} or \equ{matter-s-pot}, describing the
self-int\-er\-ac\-tion of the matter fields. Eq. 
\equ{nonren-superpot} 
is 
the content of the {\it nonrenormalization theorem for the superpotential}.

\noindent {\bf Proof of} \equ{p=0-chir-vert}:
Let us consider the contribution to the vertex function
$\G^{\rm (rad.\ corr.)}_{A_1\cdots A_N}$, taken at zero momentum, 
of a 1PI diagram containing
$n_V$, $n_S$, resp. $n_\SB$ vertices for the vector, chiral, resp. antichiral
type. 
Since all the external legs are chiral, all the variables $\tb$ are integrated.
Before these integrations, the integrand is a function 
of the differences $\tbpp$ of the $\tb$ associated to each of the vertices since
the external momenta are set to zero (c.f. \equ{fey-gen-integrand}). 
There are at most $2(n_V+n_\SB-1)$ such independent variables.  But
the total number of $\tb$-integrations is equal to $2(n_V+n_\SB)$,
which implies a vanishing integral.

\section{Renormalization}\label{renormalisation}
The material which follows is only a summary. A more complete exposition
may be found in~\cite{ps-book} 
and in the original paper~\cite{35}. 

The renormalization program consists in showing that there exists
a quantum theory, constructed as a perturbative expansion in
$\h$, whose Green functions obey all the conditions defining 
a given classical theory. If this programs succeeds, and if the
resulting theory depends on a {\it finite} number of free physical
parameters, the theory is called renormalizable.
The theory is called anomalous if the fulfilment of some of
the conditions turns out to be impossible (see~\cite{pigsor-book}). 

These conditions have been expressed 
in Section~\ref{sym} for the super Yang-Mills theory as a set of
identities (gauge condition, Ward identities, Slavnov-Taylor identity,
etc.) which the classical action $\S$ has to fulfil. 
These functionals identities have to generalize for the 
vertex functional $\G(\vf)$.
The latter is indeed the natural object
to consider in the quantum theory. It 
generates the vertex functions, i.e. the contributions
of the 1-particle irreducible Feynman graphs to the Green
functions, amputated from their external legs. 
Let us note\footnote{See Appendix~\ref{fonctionelles} for more details
on the generating functionals.} that, in the classical limit
 $\h=0$, the vertex functional coincides with the classical action:
\eq
\G(\vf) =\S(\vf) + O(\h)\ ,
\eqn{g=s+hbar}
and, for future use, that the vertex functional $\D\cdot\G(\vf)$
corresponding to a composite field insertion coincides, in the classical
limit, with the local functional (classical field polynomial) $\D$:
\eq
\D\cdot\G(\vf) =\D(\vf) + O(\h\D)\ .
\eqn{d.g=d+hbar}

\subsection{The Infrared Problem}
A difficulty, genuine to supersymmetric gauge theories in
four-dimensional space-time, is the appearance of a pseudoscalar
field $C(x)$ (the $ \theta = 0 $ component of the gauge superfield 
\equ{gauge-s-f}
which is both massless (due non-Abelian gauge invariance) and
dimensionless. Its propagator in momentum space is of the form
$1/(k^2+i 0 )^2$. It therefore presents an infrared singularity
since it is non-integrable at $k=0$. There are two known ways
out of this difficulty. The first, better known, way is to work
in the Wess-Zumino gauge~\cite{31} (see Subsection \ref{jauge de wz}), 
where the field $C$ is absent.

The second procedure~\cite{ps-book,33} for circumventing the infrared
problem consists in the introduction of a mass $\m ^2$ for the
field $C$, in such a way that the physical quantities do not
depend on $\m ^2$. This is achieved by using the possibility of
performing a non-linear field redefinition of the gauge
superfield as in \equ{8.18}, \equ{8.19}, but in a supersymmetry breaking way:
\eq
\f' = (1 + \dfrac{\m ^2}{2} \theta ^2 \bar{\th} ^2)\f
\eqn{9.1}
The propagator of $C$ becomes proportional to $1/(k^2-\m^2+
i 0)^2 \ :\  \m^2$ plays the role of an infra-red regulator.
On the other hand $\m^2$, like the parameters $a_k$ in \equ{8.19},
is a gauge parameter: the physical quantities are independent
of it. In particular the breakdown of supersymmetry,
parametrized by $\m ^2$, does not affect the physical quantities.

For details we refer the reader to the original
literature~\cite{33,ps-book}.
In these notes we shall simply assume that all fields are made
massive by adding suitable supersymmetric mass terms in the action.
Since these masses in general will break BRS invariance 
and $R$-invariance, we assume the corresponding Slavnov-Taylor and
Ward identities to hold in the asymptotic region of momentum
space only
, where the effect of the masses is negligeable.
All equalities in the
following have to be understood in this sense.

\subsection{Renormalization of the Linear Identities}
As we have outlined at the beginning of the present section,
our aim is to establish the validity, to all orders, 
of the functional identities used to define the zeroth order theory 
given by the classical action.
These identities, now written for the vertex functional $\G$
(see \equ{g=s+hbar}), are:

\noindent -- the Ward
identities for $R$-invariance, supersymmetry and rigid invariance,
\equ{sym-w-i} (translation invariance being obvious) 
\eq
W_X\G:= -i \dsum{\vf}{} \dint \d_X\vf\dfud{}{\vf}\G=0\ ,
\quad X=  R,\, Q_\a \mbox{ and rigid transf.}   \ ,
\eqn{ren-WI}
-- the gauge condition \equ{8.8} and the ghost 
equation\footnote{As we have already, said the ghost equation 
follows in fact from the gauge condition and from the Slavnov-Taylor
identity. But it is useful to begin by showing its validity, prior to
the proof of the Slavnov-Taylor identity, because it will give a further 
constraint on the possible breakings of the latter.}  \equ{8.9} 
\eq\ba{l}
{\dfrac{\d \G}{\d B}} = {\dfrac{1}{8}} \bar{D}^2 D^2 \f \ ,\es
\GG_+ \G := \lp{\dfrac{\d}{\d c_-}} + {\dfrac{1}{8}} \bar{D}^2 D^2
{\dfrac{\d}{\d \f^*}}\rp\G = 0 \ ,
\ea\eqn{ren-gauge-ghost-eq}
-- the Slavnov-Taylor identity \equ{g-class-slavnov}
\eq
\SS(\G) := \tr \iv {\dfrac{\d \G}{\d \f^*}} {\dfrac{\d \G}{\d \f}}
+ \lp \is \left\{ {\dfrac{\d \G}{\d A^{*i}}} {\dfrac{\d \G}{\d A_i}}
+ \tr {\dfrac{\d \G}{\d c_+^*}} {\dfrac{\d \G}{\d c_+}} + \tr B
{\dfrac{\d \G}{\d c_-}} \right\} + {\rm c.c.} \rp = 0 \ .
\eqn{ren-slavnov}

We begin by giving a very short description of the way 
the linear identities \equ{ren-WI}-\equ{ren-gauge-ghost-eq} may be
proven\footnote{The proof actually
given in the literature (see~\cite{ps-book}) does not take all the 
identities at once together, but treats them in sequence,
each one after the other. The present description 
(c.f.~\cite{pigsor-book}) is more concise, but equivalent.}, 
leaving the Slavnov-Taylor
identity  \equ{ren-slavnov} for the next subsection. 
Let us rewrite
the identities \equ{ren-WI}-\equ{ren-gauge-ghost-eq}, to be proven, as
\eq
\FF_A\G =0\ ,
\eqn{lin-id}
where the index $A$ enumerates all the components of each of them. We also
include the translation operators. 
The operators $\FF_A$ form a superalgebra
\eq
[\FF_A,\FF_B] = c_{ABC}\FF_C\ ,
\eqn{lin-id-alg}
-- the brackets $[\ ,\ ]$ being  commutators or anticommutators -- 
which is a
subalgebra of the complete (including BRS) algebra \equ{alg-ward-slavnov}.

The proof of the functional identities \equ{lin-id} is 
inductive and begins with the assumption that they
hold up to the loop order $n-1$:
\eq
\FF_A\G = \h^n\D_A + O(\h^{n+1})\ .
\eqn{lin-id-qap}
Due to the quantum action principle~\cite{qap},
the possible breaking in the right-hand side is 
a local field insertion,
integrated or not according to the nature of the left-hand side, and of 
dimension bounded from above by the dimension of the left-hand 
side\footnote{In fact we only take under 
consideration the terms of maximum
dimension, since the lower dimension ones mix with the breakings due to
the possible noninvariant mass terms, whose effects we have decided not
to worry with.}. 
At its lowest nonvanishing  order, i.e. at the order $n$, it is a
classical local functional $\D_A$ of the fields.

The algebraic relations \equ{lin-id-alg} applied to the vertex
functional $\G$ yield, at the order $n$, the consistency conditions
\eq
\FF_A\D_B \mp \FF_B\D_A = c_{ABC}\D_C\ .
\eqn{rig-cons}
It can be checked, in our case, 
that the general solution of the consistency conditions has the
``trivial'' form 
\eq
\D_A = \FF_A\hat\D \ ,
\eqn{rig-cons-sol}
where $\hat\D$ is an integrated local functional of dimension 4, i.e. of
the dimension of the action. We then redefine the action as
\eq
\S' = \S - \h^n\hat\D\ ,
\eqn{redef-act}
which amounts to redefine the vertex functional as
\eq
\G' = \G - \h^n\hat\D + O(\h^{n+1}) \ .
\eqn{redef-vert}
Clearly, the new vertex functional obeys the functional identities
\equ{lin-id} to the next order:
\eq
\FF_A\G = O(\h^{n+1})\ .
\eqn{next-order}
This ends the inductive proof of their validity to all orders.
\remark
The proof we have sketched includes in particular
that of the absence of anomaly for
supersymmetry, which can be found in detail 
in \cite{susy-consist}. The proof in the latter reference holds for
supersymmetric theories with a field content corresponding to the class
of super Yang-Mills theories considered here.
More general cases, where supersymmetry anomalies could occur --
although no concrete example of this is known -- were considered in 
Refs.~\cite{dixon,brandt}.
\kramer

\subsection{Renormalization of BRS Invariance}\label{ren-de-brs}
The treatment of the renormalization problem for BRS invariance, namely
the proof of the Slavnov-Taylor identity \equ{ren-slavnov}
-- with possible anomalies -- is
closely parallel to the one for the non\-sup\-er\-sym\-met\-ric 
gauge theories
discussed e.g. in~\cite{pigsor-book}. 
There is also here one single possible anomaly, which
is a supersymmetric extension of the usual Adler-Bardeen anomaly. 
It has the form of an infinite power series in the gauge
superfield $\f$:
\eq
\AA= \tr \dint d V \left( c_+ D^\a \f \bar{D}^2 D_\a \f  - 
  \bar{c}_+ \bar{D}_{\dot \a} \f D^2 \bar{D}^{\dot \a} \f + 
   O(\f^3) \right)
\eqn{9.2}
There is no simple closed expression for $\AA$ (see~\cite{34}). The
references~\cite{35,ps-book} state its existence and uniqueness. 
Explicit constructions may be found in~\cite{36}.

Let us sketch the demonstration, which makes use of the same inductive
procedure as for the linear functional identities in the last
subsection. First, through the quantum action principle and from 
the assumption that the Slavnov-Taylor identity \equ{ren-slavnov} has
been proven up to order $n-1$ in $\h$, we can write
\eq
\SS(\G) = \h^n \D\cdot\G = \h^n \D + O(\h^{n+1})\ ,
\eqn{slavnov-n}
where $\D$ is an integrated local functional of the fields, of 
dimension\footnote{Again we neglect lower dimension terms.} 4 
and ghost number 1 (see Table~\ref{table1} in Section~\ref{sym}
for dimensions and quantum
numbers). From the algebra \equ{alg-ward-slavnov} and the
fulfilment of the linear functional identities 
\equ{ren-WI}-\equ{ren-gauge-ghost-eq}, 
we deduce that the most general form for the breaking $D$ is 
restricted by the constraints 
\eq\ba{l}
W^R\D=0\ ,\quad W^Q_\a\D=0\ ,\quad W_{\rm rig}\D=0\ ,\es
\dfud{}{B}\D =0\ ,\quad \GG_+\D=0\ ,
\ea\eqn{constr-delta}
and
\eq
\SS_\S\D= 0\ ,
\eqn{brs-cons}
where $\SS_\S$ is the linearized Slavnov-Taylor operator
\equ{lin-slavnov}, $\S$ being the classical action. Due to the last of
algebra relations~\equ{alg-ward-slavnov} and to
the fulfilment of the Slavnov-Taylor
identity by the classical action, $\SS_\S$ is nilpotent:
\[
\SS_\S^2=0\ .
\]
 Solving 
\equ{brs-cons} is thus a problem of cohomology in the space of the local
functional of dimension 4, ghost number 1 and subjected to the
constraints \equ{constr-delta}. A detailed analysis shows that the BRS
constraint \equ{brs-cons} has the general solution 
\eq
\D = \SS_\S\tilde\D + r\AA\ .
\eqn{gen-sol-brs-cons}
$\tilde\D$ is an integrated local functional of dimension 4 and
ghost number 0: its absorption as a conterterm $-\h^n\tilde\D$ in the
action eliminates it from the breaking $\D$, in the same way as the
possible breakings of the linear functional identities were eliminated 
(c.f. Eqs.~\equ{rig-cons-sol} to \equ{next-order}). We are left with the
term $r\AA$, with $\AA$ given by \equ{9.2} and $r$ a calculable function of
the parameters of the theory. Since it cannot be written  
as a $\SS_\S$-variation and it represents the cohomology of the
nipotent operator $\SS_\S$ in the space of functionals under
consideration. 
From the physical point of view, $\AA$ 
represents the gauge anomaly, i.e. an
obstruction to the implementation of BRS invariance beyond the
classical approximation.  
\remarks
\item[1.]
At the one-loop order, the anomaly coefficient $r$ appears as an
algebraic expression which is the same as in the usual gauge 
theories~\cite{pigsor-book}. It follows that the absence of the anomaly
in the one-loop order is assured by the usual conditions on the choice of
the group representations for the matter fields. Its absence to all
higher orders is then assured by a supersymmetric generalization of the 
nonrenormalization theorem of Bardeen (see~\cite{pigsor-book}, e.g.). 
Although such
a generalization has not been explicitly checked, one may expect
its validity, the supersymmetric adaptation of the proof looking
obvious.
\item[2.]
The anomaly \equ{9.2} obeys the constraint
\[
\lp \is \dfud{}{c_+} + \isb \dfud{}{\cb_+} \rp \AA= 0\ ,
\]
which follows from the validity of the ``antighost equation''
\equ{g-minus} -- to be shown in Subsection~\ref{eq-de-l'antigh} --
and from the algebraic identity \equ{antic-slavnov-antigh} 
together with rigid invariance, the independence from the Lagrange
multiplyer field $B$ being taken into account (see \equ{constr-delta}).
\skramer

\subsection{The Antighost Equation}\label{eq-de-l'antigh}
It is known~\cite{bps-antigh} that in the Landau gauge -- and in some
noncovariant linear gauges as well~\cite{noncov-gauge} -- the coupling
of the Faddeev-Popov ghost $c_+$ is severely
constrained by a functional identity, the ``antighost
equation''. Its main consequence is  the nonrenormalization of the ghost
field, a property which turns out to be very useful in the proof of
various nonrenormalization theorems~\cite{lu-pi-so-BF,pi-so-ABtheor}.
Let us show that such an identity 
also holds~\cite{pig-sor-antghost}
in SYM theories in the supersymmetric Landau gauge \equ{gauge-fix}.

Differentiating the classical action \equ{tot-action} with respect to
the ghost field $c_+$ we obtain
\eq\ba{l}
\dfud{\S}{c_+} = \dfrac{1}{16} \DB^2 [D^2c_-,\f] +
   \dfrac{1}{16}\DB^2 [\DB^2\cb_-,\f] \es \phantom{\dfud{\S}{c_+}=}
  - \dfrac{1}{2} \DB^2[\f^*,\f] - \DB^2 \lp{\hat\f}^* M(\f)\rp + [c_+^*,c_+]
  +  A^*T_a A\tau_a \ ,
\ea\eqn{deriv-c}
where $M(\f)$, which appears in the nonlinear part of the BRS 
transformation of the 
gauge superfield (see \equ{brs-transf}), is defined by 
\[\ba{l} 
s\f = \half\lc\f,c_++\cb_+\rc + M(\f)(c_+-\cb_+)\ ,\es
\lp M(\f)\rp^a_b (c_+-\cb_+)^b := 
  \dfrac{1}{2}\Lp (L_\f{\rm coth}(L_\f/2))(c_+-\cb_+) \Rp^a
\ea\]

At this point one should observe that the right-hand side of \equ{deriv-c}
thus contains terms nonlinear in the quantum fields. 
These composite terms, being subject to 
renormalization, spoil the usefulness of this equation. However, considering 
the corresponding equation for $\cb_+$:
\eq\ba{l}
\dfud{\S}{\cb_+} = \dfrac{1}{16} D^2 [\DB^2\cb_-,\f] +
   \dfrac{1}{16}D^2 [D^2c_-,\f] \es \phantom{\dfud{\S}{\cb_+}=}
  - \dfrac{1}{2} D^2[\f^*,\f] + D^2 \lp{\hat\f}^* M(\f)\rp + [\cb_+^*,\cb_+]
  -  \AB^*T_a \AB\tau_a \ ,
\ea\eqn{deriv-cbar}
adding together the superspace integrals of the equations 
\equ{deriv-c}, \equ{deriv-cbar} and using\footnote{Use has to be made 
of the identity
\[
\is\DB^2[D^2c_-,\f] = \is[c_-, \DB^2D^2\f]\ ,
\]
and of its complex conjugate.}
 the Landau gauge condition \equ{8.8},
one obtains the antighost equation we are looking for:
\eq
\GG_-\S = \D_{\rm class}\ ,
\eqn{class-antig-eq}
with
\eq
\GG_-:= \is \lp \dfud{}{c_+} -\lc c_-,\dfud{}{B} \rc\rp
  + \isb \lp \dfud{}{\cb_+} -\lc \cb_-,\dfud{}{\BBAR} \rc\rp
\eqn{g-minus}
and
\eq
\D_{\rm class} := -\iv [\f^*,\f]
  + \is\lp [c_+^*,c_+] + (A^*T_a A)\tau_a \rp
  + \isb\lp [\cb_+^*,\cb_+] - (\AB T_a \AB^*)\tau_a \rp\ .
\eqn{class-break}
We remark that the undesired nonlinear terms present in each of 
the equations \equ{deriv-c} and \equ{deriv-cbar} have been cancelled. 
We are thus left with the breaking \equ{class-break} which, 
being now linear in the quantum fields, will not be renormalized, 
i.e., it will remain a classical breaking.

Equation \equ{class-antig-eq} has now a form which allows one 
to consider its validity to all orders of perturbation theory. 
That it indeed holds as it stands at the quantum level:
\eq
\GG_-\G = \D_{\rm class}\ ,
\eqn{antig-eq}
 may be shown without any difficulty by repeating exactly the 
argument given in~\cite{bps-antigh,pigsor-book} 
for the nonsupersymmetric case.

Let us finally remark that the sum of the superspace-integrated functional 
derivatives with respect to $c_+$ and $\cb_+$ in \equ{g-minus} is in 
fact the space-time integral of the functional derivative with respect 
to the real part of the $\theta=0$ component of $c_+$. It coincides with 
the functional operator appearing in the nonsupersymmetric 
version of the antighost equation. 

We shall now derive an interesting consequence of the antighost equation.
Using the ``anticommutation relation''
\eq
\GG_-\SS(\G) + \SS_\G\lp \GG_-\G - \D_{\rm class}\rp = W_{\rm rig}\G\ ,
\eqn{antic-slavnov-antigh}
where $\SS_\G$ is the linearized Slavnov-Taylor operator defined
accordingly to \equ{lin-slavnov}, and
\eq\ba{l}
W_{\rm rig}\G := \iv\lp \lc \f,\dfud{\G}{\f}\rc 
   + \lac \f^*,\dfud{\G}{\f^*}\rac\rp \es
+\is\lp \lac c_+,\dfud{\G}{c_+}\rac + \lc c^*_+,\dfud{\G}{c^*_+}\rc
      + \lc B,\dfud{\G}{B}\rc + \lac c_-,\dfud{\G}{c_-}\rac 
      +\lp\dfud{\G}{A} T_a A\rp\tau_a 
             - \lp A^* T_a \dfud{\G}{A^*}\rp\tau_a \rp \es
+\isb\lp \lac \cb_+,\dfud{\G}{\cb_+}\rac + \lc \cb^*_+,\dfud{\G}{\cb^*_+}\rc
      + \lc \BBAR,\dfud{\G}{\BBAR}\rc + \lac \cb_-,\dfud{\G}{\cb_-}\rac 
      -\lp\AB T_a\dfud{\G}{\AB}\rp\tau_a  
        - \lp \dfud{\G}{\AB^*} T_a \AB^* \rp\tau_a \rp \es
 = 0 \ .
\ea\eqn{wi-rig}
One thus sees that, in the Landau gauge, the identity
\eq
W_{\rm rig}\G=0
\eqn{rig-wi}
follows from the Slavnov-Taylor identity and the antighost equation. 
This is the Ward identity expressing the invariance of the theory
under the rigid transformations \equ{rig-transf}, corresponding to the 
transformations of the gauge group with constant parameters.

\subsection{Invariant Counterterms}\label{contretermes-inv}
Once the gauge fixing condition \equ{8.8}, the ghost equation\equ{8.9}, 
the Slavnov-Taylor identity \equ{g-class-slavnov}, 
the Ward identity for $R$-invariance
(third of equs. \equ{sym-w-i}) and 
the antighost equation \equ{antig-eq}
have been established\footnote{Poincar\'e invariance and supersymmetry
are obvious since the renormalization scheme 
preserves them explicitly.}
at a given arbitrary order $\h^n$ as shown in Section \ref{renormalisation}, we are still free to introduce 
in the action, at the same order, 
counterterms which do not spoil these
identities. A generic ``invariant counterterms'' $\D$ has thus to
obey to the constraints
\eq\ba{l}
\dfud{\D}{B} = 0\ ,\quad \GG_+\D = 0\ , \quad
W_R\D = 0 \ ,\quad W_{\rm rig}\G=0 \ ,\quad \GG_-\D = 0\ ,\es
\SS_\S\D = 0\ ,
\ea\eqn{invar-c-terms}
where $\SS_\S$ is the linearized Slavnov-Taylor operator
\equ{lin-slavnov}. 
The ghost number of $\D$ is 0 and its dimension 4 \footnote{We 
don't consider counterterms of lower dimension, as they would affect the
mass terms which anyhow break these symmetries.}.

The general solution of these constraints is a linear superposition of
the following terms:
\eq\ba{l}
\tr\is F^\a F_\a\ ,\quad \is\la_{(ijk)}A^iA^jA^k + {\rm c.c.}\ ,\quad
  \SS_\S \tr\is A^*_i A^j + {\rm c.c.}\ ,\es
\SS_\S \tr\iv {\hat\f}^*\f\ ,\quad \SS_\S \tr\iv {\hat\f}^*(\f)^k\ ,
\ea\eqn{class-c-terms}
where $\S$ is the most general classical action as given by \equ{8.16},
and ${\hat\f}^*$ is the shifted external superfield \equ{8.11}.

Another, more convenient, basis for the invariant counterterms is given
by the expressions
\eq
\nabla_I \S\ ,
\eqn{class-c-terms'}
where the operators $\nabla_I$ respectively are
\eq\ba{l}
\pa_g\ ,\ \pa_{\lambda_{ijk}}\ ,\ \NN_\f = N_\f - N_{\f^*} -N_{c_-} -
N_{\bar{c}_-}- N_B - N_{\bar{B}}\ ,    \es
\NN^j_i = \dint d S \left( A^j \dfrac{\delta}{\delta A^i} - A^*_i
\dfrac {\delta}{\delta A^*_j} \right) + {\rm conj.} \ ,    \es
\pa_{a_k} \ , 
\ea\eqn{10.23}
where we have introduced the ``counting operators'' 
\eq
N_\vf = \dint \vf \dfrac{\d}{\d \vf}\ ,\quad 
  \vf= \f,\, \f^*,\, c_\pm,\, B\ .
\eqn{10.9}
The invariance of the expressions \equ{class-c-terms'} follows from the
operators \equ{10.23} being ``symmetric'', i.e. 
from their ``(anti)commutativity''\footnote{There are in fact
two nonvanishing anticommutators, namely
\[
\lc \NN_\f, \dfud{}{B} \rc = \dfud{}{B}\ ,\quad
\lc \NN_\f, \GG_+ \rc = \GG_+\ ,
\]
but this has no consequence since they are applied to the action which
obeys the gauge condition and the ghost equation.} with
the operators appearing in the constraints \equ{invar-c-terms}:
\eq\ba{l}
\nabla_I \SS(\g) - \SS_\g \nabla_I \g = 0 \ ,\quad\forall \g\ ,\es
\lc\nabla_I,\dfud{}{B}\rc = 0\ ,\quad
\lc\nabla_I,\GG_\pm\rc = 0\ ,\quad \lc\nabla_I,W_R\rc = 0\ .
\ea\eqn{sym-op}
This is clear for $\pa_g$, $\pa_\la$ and $\pa_{a_k}$. For the operators
$\NN$ this follows from
\[
\NN_\f \g=\SS_\g\tr\iv{\hat\f}^*\f\ ,\quad 
\NN_i^j \g=\SS_\g\tr\is A_i^*A^j + {\rm conj.}\ .
\]
One thus sees that the counterterms, in the form
\equ{class-c-terms'}, correspond to a renormalization of the
parameters of the action and of the field amplitudes.
This shows the stability of the theory under the perturbative quantum
fluctuations.
\remarks
\item[1.]
The latter property, which is equivalent to the stability of the classical
action under the effect of small perturbations, characterizes the
renormalizability of the theory.
\item[2.]
A renormalization of the ghost field $c_+$ would be implied by a
counterterm 
\[\ba{l}
-\SS_\S \tr\is c^*_+ c_+  + {\rm conj.} = 
   \lp  N_{c_+}-N_{c^*_+} + {\rm conj.} \rp\S \es
= \tr\is c_+^* sc_+  + {\rm conj.} + \is A^*sA + {\rm conj.}
  + \iv {\hat\f}^*s\f   \ .
\ea\]
But such a term, depending on the superfield $c_+$ without derivative,
is forbidden by the last of the constraints \equ{invar-c-terms},
which corresponds to the antighost equation \equ{antig-eq}.
\skramer
The coefficients of the invariant counterterms \equ{class-c-terms}
or \equ{class-c-terms'} are still
free parameters. They are usually fixed by imposing normalization
conditions which define the field amplitudes and the physical parameters
of the theory~\cite{pigsor-book}. In our case we can choose the conditions
given in Eqs. (5.180-182) of Ref.~\cite{ps-book}. We only mention here
that they involve vertex functions taken at some fixed 4-impulsions
characterized by a normalization mass $\k$, and that, in the tree
approximation, they reproduce the parametrization of the classical
action given in section \ref{sym}.

\subsection{Callan-Symanzik Equation}
The classical theory is scale invariant if all the fields are massless,
or at least asymptotically scale invariant if 
there are masses. This is no longer true for
the quantum theory. This ``scale anomaly'' is best expressed by the
Callan-Symanzik equation. In order to derive it, let us introduce the 
dilatation generator 
\eq
D := \dsum{{\rm all\ mass\ parameters\ } m}{} m\dpad{}{m}\ ,
\eqn{dil-op}
where the summation is taken over all the mass parameters, 
including the normalization mass $\k$ introduced through the 
normalization
conditions. Application of this operator to the classical action yields
the equation 
\eq
D\S = 0\ ,
\eqn{class-CS}
which expresses the
asymptotic  scale invariance of the classical 
theory\footnote{Recall that (non-gauge invariant) mass terms being
implicitly present, such an  equality is valid up to terms of dimension 
less than the dimension of
the left-hand-side, i.e. less than 4. These terms are negligible at high
momenta.}.

Let us now apply the dilatation generator to the full vertex functional
$\G$. Through the quantum action principle, we obtain
\eq
D\G = \D\cdot\G\ ,
\eqn{dg=deltag}
where $\D$ is an insertion of dimension 4, of order $\h$ and
whose effect is to break asymptotic scale invariance. 

Noting that $D$ is a symmetric operator according to the definition
given in \equ{sym-op}, we conclude that $\D$ is an invariant insertion,
which we can expand in a basis of invariant dimension 4 insertions. 
Such a basis may be provided by the set of insertions
\eq
\lac \nabla_I \G \rac\ ,
\eqn{quantum-c-terms}
where the $\nabla_I$'s are the symmetric operators \equ{10.23}. This
is a quantum extension of the classical basis of
counterterms \equ{class-c-terms'}. The expansion of the right-hand side of 
\equ{dg=deltag} yields the Callan-Symanzik equation
\eq
C \Gamma := \left( D + \b_g \pa_g + \b_{ijk}
\pa_{\lambda_{ijk}} - \gamma_\f \NN_\f - \gamma^i_j \NN^j_i 
- \gamma_k \pa_{a_k}  \right) \Gamma = 0 \ . 
\eqn{10.25}
The coefficients $\b$ and $\g$ are of order $\h$. The former
correspond to the renormalization of the coupling constants, the latter
-- the ``anomalous dimensions'' -- to the renormalization of the field
amplitudes and of the unphysical parameters $a_k$ 
(see \equ{8.19}-\equ{8.20}).
\remark 
There is no anomalous dimension for the ghost field $c_+$. This is a
consequence of the antighost equation \equ{antig-eq} (see the second
remark at the end of Subsection~\ref{contretermes-inv}).
\kramer 



\newcommand{\wh}{{\hat w}} \newcommand{\WH}{{\hat W}}
\newcommand{\wb}{{\bar w}} 
\newpage
\section{Supercurrent}
The matter presented 
in this section is extracted from the original papers~\cite{39}, the
book~\cite{ps-book}, with slight modifications introduced later
in the papers~\cite{ps-fin-ijmp,lps-fin-hpa}.
\subsection{Classical Theory}
We have seen that supersymmetry and BRS invariance
together with power counting fix the action \equ{8.16}. 
 We have also seen (Subsection~\ref{invariance R}) that the latter
turns out to be also invariant\footnote{{}I recall that due to the
fields being all massive in order to avoid infrared problems, BRS
as well as $R$-invariance hold only asymptotically
in momentum space. Hence all the following
equations are meant to hold asymptotically only.}
under the following chiral phase
transformation, called $R$-invariance~\cite{37} (we ommit the
infinitesimal parameter):
\eq
\d^R \vf = i \left( n_\vf + \th^\a \dfrac{\pa}{\pa \th^\a} -
\bar{\th}^{\dot\a} \dfrac {\pa}{\pa \bar{\th}^{\dot \a}}  \right)
\vf \ .
\eqn{10.1}
The ``R-weights'' $n_\vf$ are given in table~\ref{table1} 
(Subsection~\ref{invariance R})

\noindent Let us recall that
the $R$-transformation commutes with BRS, but not with supersymmetry:
\eq
[\d^Q_\a,\d^R] = i\d^Q_\a\ ,\qquad [\d^{\bar Q}_{\dot\a},\d^R] =
- i\d^{\bar Q}_{\dot \a} \ .  
\eqn{10.2}
Taking into account the Wess-Zumino algebra \equ{alg-diff-op} 
we see that
the generators of R-trans\-form\-at\-ions, supersymmetry and
translations form a supermultiplet, supersymmetry acting on them by
(anti)commutation. The supercurrent~\cite{38} is then the supermultiplet
which contains the conserved Noether currents $R_\m , \ Q_{\m
\a}$ and $T_{\m \n}$ associated respectively to the invariances
under $R$, supersymmetry and translations. The supercurrent is
represented by a vector superfield
\eq
V_\m (x,\th,\bar{\th}) = R_\m(x) - i \th^\a Q_{\m \a} (x)
+ i \bar{\th}_{\dot \a} \bar{Q}^{\dot{\a}}_\m (x) - 2 (\th \s^\n
\bar{\th}) T_{\m \n}(x) + \cdots 
\eqn{10.3}
where we have written only the most relevant terms. $T_{\m\n}$ is the
``improved'' (i.e. symmetric, traceless in the classical
approximation) energy-momentum tensor. $Q_{\m \a}$ is also
traceless  in the classical approximation
-- in the sense: $\s^\m_{\a \dot \a} \bar{Q}^{\dot\a}_\m = 0$.

We are going to show that the precise identification of these
currents and of their properties follow from the {\it supertrace identities}, 
written here in the classical approximation: 
\eq\ba{l}
\bar{D}^{\dot \a} V_{\a \dot \a} = -2w_\a \S - \dfrac{4}{3} D_\a S^0 \ ,
\quad D^\a V_{\a\ad} = -2\bar w_\ad \S - \dfrac{4}{3} \DB_\ad\SB^0 \ ,\es
\mbox{with}\quad 
V_\m = \s^{\a \dot \a}_\m V_{\a \dot \a} \ , \quad V_{\a \dot
\a}  = \half \s_{\a \dot \a}^\m V_\m \ ,
\ea\eqn{10.5} 
where
$\S$ is the classical action \equ{tot-action},
and $w_\a$ is the functional differential
operator (we use the definitions of 
Refs.~\cite{ps-fin-ijmp,lps-fin-hpa}, slightly
different from those of~\cite{39,ps-book})\footnote{For 
a chiral field $\vf$ of weight $n$: $\quad
w_\a= n D_\a\lp\vf\dfud{}{\vf}\rp+2D_\a\vf\dfud{}{\vf}\ .$\\
For a vector superfield $V$ (of weight 0), on may have:
\[
\mbox{either:}\quad w_\a = -2\lp \bar{D}^2 D_\a V \dfrac{\d}{\d V} 
 - D_\a V \bar{D}^2 \dfrac{\d}{\d V} \rp \ ,
\quad\mbox{or:}\quad
w_\a = 2\lp V \bar{D}^2 D_\a\dfrac{\d}{\d V} 
 -  \bar{D}^2 V D_\a \dfrac{\d}{\d V} \rp \ .
\] }
\eq \ba{l}
w_\a = 2\tr \left( - \bar{D}^2 D_\a \f \dfrac{\d}{\d\f} + D_\a
\f \bar{D}^2 \dfrac{\d}{\d\f} \right. 
+  \f^* \bar{D}^2 D_\a \dfrac{\d}{\d\f^*} - \bar{D}^2 \f^* D_\a
\dfrac{\d}{\d\f^*}              \es\phantom{w_\a = 2\tr \LP}
\quad\quad + D_\a c_+ \dfrac{\d}{\d c_+} 
\left. - c_+^* D_\a \dfrac{\d}{\d c_+^*} - c_- D_\a \dfrac{\d}{\d c_-}
 - B D_\a \dfrac{\d}{\d B} \right)    \es\phantom{w_\a =}
+\dfrac{4}{3} D_\a A\dfrac{\d}{\d A} 
          - \dfrac{2}{3} A D_\a \dfrac{\d}{\d A} +
\dfrac{2}{3} D_\a A^* \dfrac{\d}{\d A^*} - \dfrac{4}{3} A^* D_\a
\dfrac{\d}{\d A^*} \ . 
\ea\eqn{10.6}
This operator is BRS-symmetric, i.e.:
\[
w_\a \SS(\g) - \SS_\g w_\a = 0\ ,\quad\forall \g\ .
\]
The chiral superfield\footnote{This form of $S^0$ is the one found 
in~\cite{ps-fin-ijmp,lps-fin-hpa}. It differs of the form given 
in~\cite{39} or in~\cite{ps-book} due to a different choice for the gauge
condition.}  
$S^0$ is a polynomial in the superfields
$\f , \ c_-$ and $B$:
\eq
S^0 = \dfrac{1}{8} s\, \tr \Lp \bar{D}^2 ( 2 c_- D^2 \f - D^2 c_- \f)
    \Rp + {\rm conj.} \ ,
\eqn{10.7}
which is not gauge invariant.
It is however BRS invariant, but nonphysical since it is a $s$-variation. 
It possesses the properties
\eq\ba{l}
\dint d SS^0 - \dint d \bar{S}\SB^0  =  0    \es
\dint d SS^0 + \dint d \bar{S}\SB^0   =  2(N_B + N_{c_-} + {\rm
conj.}) \S \ ,   
\ea\eqn{10.8}
 where $N_\vf$ is the $\vf$-field counting operator \equ{10.9}.

Without going into details, (see~\cite{ps-book,39}) 
let us write the BRS invariant supercurrent, 
solution of the supertrace identities, for the classical theory: 
\eq\ba{l} 
V_{\a \dot \a}= \dfrac{1}{6} [D_\a,\bar{D}_{\dot \a}]
(\bar{A} e^{T^a \f_a} A ) 
+\half \bar{D}_{\dot \a} (\bar{A} e^{T^a \f_a}) e^{-T^a \f_a} D_\a  (e^{T^a
\f_a} A)    
+ \dfrac{1}{16} \tr \left( F^\a e^{-\f} F_\a e^\f \right) 
+ \cdots \ ,
\ea\eqn{10.4}
 with $F_\a$ given in \equ{sym-action}, and
where the dots represent non-gauge-invariant terms  
produced by the gauge fixing, ghost terms
and external field contributions. 

In order to see that the supertrace identities \equ{10.5} yield 
the conservation of the currents associated to
$R$, supersymmetry and translation invariances, let us define 
the superfield currents
\eq\ba{l}
\hat R_\m := V_\m  =: R_\m +O(\th) \ ,\es
\hat Q_{\m\a} := i\lp D_\a V_\m - (\s_\m\sb^\n D)_\a V_\n \rp
   =: Q_{\m\a} +O(\th) \ ,\es
\hat T_{\m\n} := -\dfrac{1}{16} \lp V_{\m\n} + V_{\n\m} 
  - 2g_{\m\n}{V_\la}^\la \rp
   =: T_{\m\n} +O(\th)   \ ,\es\phantom{\hat T_{\m\n}}
\mbox{with}\quad V_{\m\n} := 
  \s_{\m\b\bd} [D^\b,\DB^\bd] V_\n \ .
\ea\eqn{comp-curr}
We first check the conservation law of $\hat R_\m$, i.e. of
the supercurrent $V_\m$.
It is obtained by applying $D^\a$ on the first of the supertrace identities
\equ{10.5}, $\DB^\ad$ on the second one, then adding together
the identities thus obtained , and finally using the anticommutation 
rule \equ{alg-cov-der} for the covariant derivatives. We get 
\eq\ba{l}
\pa^\m V_\m = i\wh^R\S + \dfrac{2}{3}i(D^2S^0-\DB^2\SB^0)\ ,\es
\mbox{with}\quad \wh^R  = D^\a w_\a - \bar{D}_{\dot \a} \bar{w}^{\dot \a} \ .
\ea\eqn{s-curr-conserv}
The other conservation laws follow from the latter identity and from the 
supertrace identity. One finds, altogether:
\eq\ba{l}
\pa^\m\hat R_\m = i\wh^R\S + \dfrac{2}{3}i(D^2S^0-\DB^2\SB^0)\ ,\es
\pa^\m\hat Q_{\m\a} =i\wh^Q_\a\S \ ,\es
\pa^\m\hat T_{m\n} = i\wh^P_\n\S \ ,
\ea\eqn{comp-conserv}
with
\eq\ba{l}
\wh^R  = D^\a w_\a - \bar{D}_{\dot \a} \bar{w}^{\dot \a} \ ,\es
\wh^Q_\a = iD_\a(D^\b w_\b-\DB_\bd{\bar w}^\bd)
   - 4i\s^\m_{\a\ad} \pa_\m {\bar w}^\ad \ , \es
\wh^P_\n = -\dfrac{1}{16}\sb_\n^{\ad\a}\lp D^2\DB_\ad w_\a 
 + \DB^2D_\a\wb_\ad + [D_\a,\DB_\ad](D^\b w_\b-\DB_\bd\wb^\bd) \rp
  + \dfrac{i}{2}\pa_\n(D^\b w_\b+\DB_\bd\wb^\bd) \ .
\ea\eqn{comp-ct}
One checks that the space-time integration of the latter
functional operators at $\th=0$ yields the Ward operators of the
corresponding symmetries:
\eq
\xint \wh^R = W^R + O(\th)\ , \quad \xint \wh^Q_\a = W^Q_\a + O(\th)\ , 
 \quad \xint \wh^P_\n = W^P_\n + O(\th)\ , 
\eqn{super-ward-op}
It follows that, taken at $\th=0$, the equations 
\equ{comp-conserv} express the
conservation of the Noether currents
$R_\m$, $Q_{\m\a}$, $T_{\m\n}$  associated to $R$-invariance,
supersymmetry and translation invariance, respectively, identified as
the $\th=0$ components of the superfield 
currents \equ{comp-curr}. (The
conservation of $R_\m$ holding up to the nonphysical breaking in $S^0$). 

\noindent
One also finds that, 
beyond the conservation laws \equ{comp-conserv}, the identities
\equ{10.5} also contain the {\it trace identity}
\eq
{\hat T}_\la^\la =  -\dfrac{3}{2}\lp D^\a w_\a+\DB_\ad{\bar w}^\ad \rp\S
  - \lp D^2S^0+\DB^2\SB^0 \rp\ .
\eqn{trace-id}
Thus the conserved energy-momentum tensor contained in the 
supercurrent is symmetric and traceless
(up to the nonphysical $S^0$-terms): 
this identifies it as the improved energy-momentum
tensor. Defining thus the current
\eq
{\hat D}_\m = x^\n {\hat T}_{\m\n} =: D_\m + O(\th)\ ,
\eqn{super-dil}
we check that the latter
is conserved (modulo the nonphysical breaking terms in $S^0$):
\eq
\pa^\m {\hat D}_\m = \lp ix^\n\wh^P_\n 
  -\dfrac{3}{2}(D^\a w_\a+\DB_\ad {\bar w}^\ad) \rp    \S 
    - \lp D^2S^0+\DB^2\SB^0 \rp\ .
\eqn{dil-curr-conserv}
Its $\th=0$ component $D_\m$ is nothing else than the dilatation current,
conserved in the classical ap\-prox\-im\-at\-ion\footnote{We recall 
that we
neglect every breaking due to the masses.}. Indeed, the space-time
integration of the right-hand side of 
\equ{dil-curr-conserv} yields, with \equ{10.8} taken into account, the
dilatation Ward identity:
\eq\ba{l}
\xint \lp ix^\n\wh^P_\n 
  -\dfrac{3}{2}(D^\a w_\a+\DB_\ad {\bar w}^\ad) \rp    \S 
    - \lp \is S^0+ \isb\SB^0 \rp = iW^D\S +O(\th) = 0\ ,\es
\mbox{with}\quad 
 W^D := -i\dsum{\vf}{} \dint \d_D \vf \dfrac{\d}{\d\vf} \ ,\quad
   \d_D \vf = \left( d_\vf + x^\m \pa_\m 
  + \half \th^\a \dfrac{\pa}{\pa \th^\a} + \half{\bar{\th}^{\dot \a}}
    \dfrac{\pa}{\pa \bar{\th}^{\dot \a}} \right) \vf \ ,
\ea\eqn{10.17}
$d_\vf$ being the dimension of the superfield $\vf$.
\remarks
\item[1.]
The scale dimensions $d_\vf$ of the superfields contained in
this theory are the canonical ones. They are
given in Table~\ref{table1} (in Subsection~\ref{invariance R}).
Looking to \equ{10.8}, one sees that the $S^0$-term 
in the right-hand side of 
\equ{trace-id} does contribute to the dimensions of
$B$ and $c_-$. If it were absent, the wrong dimension 3 would have been 
obtained for these two fields.
\item[2.]
The supertrace identities \equ{10.5} also imply the ``spinor trace
identity'' 
\[
\sb_\m^{\ad\a}{\hat Q}^\m_\a = -12i{\bar w}^\ad \S - 8i\DB^\ad\SB^0\ .
\]
This allows to define (at $\th=0$9)
\[
 S_{\m\a} = ix_\n\s^\n_{\a\ad} {\bar Q}_\m^\ad\ ,
\]
the Noether current asssociated to conformal
supersymmetry.  This, together with the special
conformal current
\[
K_{\m\n} = \lp 2x_\n x^\la-\d^\la_\n x^2 \rp T_{\m\la}\ ,
\]
completes the list  of the Noether currents 
associated to the superconformal group~\cite{wz-first,38,ps-book}.
\skramer

\subsection{Renormalization of the Supercurrent}
\subsubsection*{Statement of the Result}
One has to show that the supercurrent identities \equ{10.5} are
renormalizable~\cite{ps-book,39} in the sense
that there exists a BRS-invariant quantum extension of the
supercurrent \equ{10.4} and a chiral insertion $S$ of dimension 3
and R-weight -2, such that the identities
\eq\ba{l}
\bar{D}^{\dot \a} V_{\a \dot \a} \Gamma = -2 w_\a \Gamma - \dfrac
{4}{3} D_\a (S + S^0)\cdot \Gamma \ ,\es
 D^\a V_{\a\ad} \Gamma = -2 \bar w_\ad \Gamma - \dfrac
{4}{3} \DB_\ad (\SB + \SB^0)\cdot \Gamma \ ,
\ea\eqn{10.14}
hold to all orders. $S^0$ is now a quantum extension of \equ{10.7},
which will be defined later on
in such a way that it remains a BRS variation -- i.e., now,
a variation under the 
linearized Slavnov-Taylor operator $\SS_\G$ -- and that it 
still obeys the identities \equ{10.8}.
The new chiral insertion $S\cdot\G$ in \equ{10.14} is an anomaly, which
does not spoil the conservation of the currents
$Q_{\m\a}$ and $T_{\m\n}$ defined from $V_\m$ by \equ{comp-curr}, 
but breaks\footnote{Recall that the
term in $:S^0:$ and conj. does not represent a breaking, 
since it is a total derivative (see the first of Eqs. \equ{10.8}), and
moreover is nonphysical, being a BRS varition.} 
the conservation of 
$\hat R_\m$ (c.f. \equ{comp-conserv}):
\eq
\pa_\m \hat R^\m \cdot \Gamma = i\left. \wh^R \Gamma \right|_{\th = 0} 
  + \dfrac{2}{3}i \lp D^2 S^0 - \bar{D}^2 \SB^0 \rp\cdot \Gamma 
  + \dfrac{2}{3}i \lp D^2 S - \bar{D}^2 \SB \rp\cdot \Gamma \ , 
\eqn{10.15}
and similarly gives an anomaly to the trace identity \equ{trace-id}:
\eq \ba{l}
{\hat T}_\la^\la\cdot\G =  
 -\dfrac{3}{2}\lp D^\a w_\a+\DB_\ad{\bar w}^\ad \rp\G
  - \lp D^2S^0+\DB^2\SB^0 \rp\cdot\G 
  - \lp D^2S+\DB^2\S \rp\cdot\G \ .
\ea\eqn{10.16}
The space-time integral of the
sum of the first two terms in the right-hand-side being equal,
at $\th=0$, to the action on $\Gamma$ 
of the dilatation Ward identity operator \equ{10.17}, this yields, as
in the classical limit, the dilatation Ward identity, 
but now broken by the scale anomaly $S$:
\eq
W_D \Gamma = -i \left( \dint dSS + \dint d \bar{S} \bar {S} \right) 
\cdot\Gamma \ . 
\eqn{10.20}
The scale anomaly in the r.h.s. of \equ{10.20} can be written in a
suggestive way by expanding the dimension 3 chiral insertion
$S$ in an appropriate BRS invariant basis: 
\eq
S = \b_g L_g + \dsum{}{} \b_{ijk} L_{ijk} - \gamma_\f L_\f - \dsum{}{}
\gamma^i_j L^j_i  - \dsum{k}{} \gamma_k L_k \ . 
\eqn{10.21}
The basic elements $L_I$ are defined (up to total derivatives)
through the action principle by
\eq
\left( \dint d S L_I + \dint d \bar{S} \bar{L}_I \right)\cdot \Gamma =
\nabla_I \Gamma \ , 
\eqn{10.22}
whith the ``symmetric operators'' $\nabla_I$  defined by
\equ{10.23}. Using now the dimension analysis identity
\eq
iW^D \Gamma + D \Gamma = 0 \ , 
\eqn{10.24}
 where $D$ is the dilatation generator \equ{dil-op},
we see that the broken Ward identity \equ{10.20} is nothing else than
the Callan-Symanzik equation \equ{10.25}.
\subsubsection*{Sketch of the Proof of the Renormalized Supertrace
Identities} 
The quantum action principle yields
\eq
w_\a\G = \D_\a\cdot\G\ ,
\eqn{qap-w-alpha}
where $\D_\a$ is an insertion of dimension 7/2. Due to the algebraic
identities 
\eq\ba{l}
w_\a\SS(\g) - \SS_\g w_\a\g = 0\ ,\quad\forall\,\g\ ,\es
[W^R,w_\a] = \lp-1+\th\dpad{}{\th}+\tb\dpad{}{\tb}\rp w_\a  \ ,\es
\lc\dfud{}{B(1)},w_\a(2)\rc = -2\d_S(1,2) D_\a\dfud{}{B(2)}
             \ ,\quad \lc\dfud{}{\BBAR(1)},w_\a(2)\rc = 0\ ,\\[5mm]
\lp\GG_+(1),w_\a(2)\rp = \es\quad
-2\d_S(1,2) D_\a\dfud{}{c_-(2)}
  + \dfrac{1}{4}\DB^2D^2\d_V(1,2)\DB^2D_\a\dfud{}{\f^*(2)}
  - \dfrac{1}{4}\DB^2D^2\DB^2\d_V(1,2)D_\a\dfud{}{\f^*(2)}\ ,\es
\lp\bar\GG_+(1),w_\a(2)\rp = 
    \dfrac{1}{4}D^2\DB^2\d_V(1,2)\DB^2D_\a\dfud{}{\f^*(2)}\ ,
\ea\eqn{comm-w-alpha}
the insertion $\D_\a$ is submitted to the constraints
\eq\ba{l}
\SS_\G\D_\a\cdot\G = 0 \ ,\es
W^R\D_\a\cdot\G = \lp-1+\th\dpad{}{\th}+\tb\dpad{}{\tb}\rp \D_\a\cdot\G \ ,\es
\dfud{}{B(1)} \D_\a(2)\cdot\G = \dfrac{1}{4}\lp \DB^2D^2\d_S(2,1)D_\a\f(2)
  - D^2\d_S(2,1)\DB^2D_\a\f(2) - \d_S(2,1)D_\a\DB^2 D^2\f(2) \rp\ ,\es
\dfud{}{\BBAR_(1)} \D_\a(2)\cdot\G = 
  - \dfrac{1}{4} \DB^2\d_\SB(2,1)\DB^2D_\a\f(2)  \ ,\\[5mm]
\GG_+(1) \D_\a(2)\cdot\G  \es\quad
 = - \dfrac{1}{4}\lp \DB^2D^2\d_S(2,1)D_\a \,s\f(2)
  - D^2\d_S(2,1)\DB^2D_\a \,s\f(2) 
  - \d_S(2,1)D_\a\DB^2 D^2 \,s\f(2) \rp       \ ,\es
\bar\GG_+(2) \D_\a(2)\cdot\G =   
    \dfrac{1}{4} \DB^2\d_\SB(2,1)\DB^2D_\a \,s\f(2)  \ .
\ea\eqn{d-alpha-constraints}
We have to solve these constraints. 
Let us first look for a special solution.
Such a solution is given, at the classical level, by the 
expression\footnote{\equ{delta-alpha-0} being in fact obtained 
through applying the operator $w_\a$ \equ{10.6} on the gauge fixing term 
\equ{gauge-fix} of the classical action, the fulfilment of the constraints
\equ{d-alpha-constraints} is obvious.}
\eq
\D_\a^0 = -\half\DD^\ad V^0_{\a\ad} -\dfrac{2}{3}D_\a S^0\ ,\es
\eqn{delta-alpha-0}
with
\eq\ba{l}
\D_\a^0 = s\,\hat\D_\a\ ,\quad V^0_{\a\ad} = s\,\hat V_{\a\ad} \  , 
   \quad S^0 = s\,\hat S\ ,\es
\quad \hat\D_\a = \dfrac{1}{4}\tr\lp \DB^2D^2c_-D_\a\f - D^2c_-\DB^2D_\a\f
  - c_-D_\a\DB^2D^2\f - \DB^2\cb_-\DB^2D_\a\f \rp \ ,   \es
\quad \hat V_{\a\ad} = \dfrac{1}{3}\tr\Lp 
 D_\a\DB_\ad D^2c_-\f + \DB_\ad D^2c_-D_\a\f - \DB_\ad D_\a c_-D^2\f \es
\phantom{\quad \hat V_{\a\ad} = \dfrac{1}{3}\tr\Lp} 
 + D^2c_-[D_\a,DB_\ad]\f - D_\a c_-\DB_\ad D^2\f + c_-D_\a\DB_\ad D^2\f 
 \Rp\ +\ \mbox{conj.} \ ,\es
\quad \hat S = \dfrac{1}{8} \tr \lp \bar{D}^2 ( 2 c_- D^2 \f - D^2 c_- \f)
    \rp + {\rm conj.} \ .
\ea\eqn{hat-d-alpha-0}
(The expression for $S^0$ is the same as given in \equ{10.7}.) 
We notice that the ``hat'' quantities $\hat\D_\a$, etc.,
obey the same identity 
\equ{delta-alpha-0} as their BRS variations $\D^0_\a$, etc.

We shall define the quantum extensions of the expressions $\D_\a^0$, etc., 
as the quantum BRS variations -- i.e. the variations under $\SS_\G$ -- 
of the ``Wick products'' of the bilinear expressions $\hat\D_\a$, etc.:
\eq
\D_\a^0 = \SS_\G :\hat\D_\a:\ ,\quad V^0_{\a\ad} = \SS_\G :\hat V_{\a\ad}: \  , 
   \quad S^0 = \SS_\G :\hat S:\ .
\eqn{q-delta-alpha-0}
We define
the Wick product of a bilinear expression $AB$ at a superspace point
$(x,\th)$ as the insertion $:AB:$ obtained by subtracting off
the infinite part of the Wilson expansion~\cite{wilson-zim} of the
bilocal $T$-product $T(A(x+\e,\th)B(x-\e,\th))$ at $\e^\m=0$:
\[
\mbox{Fin. part }\lim_{\e\to0} T\lp A(x+\e,\th)B(x-\e,\th)\rp =
 \, :AB:(x,\th) \ .
\]
Since these renormalized quantities obey the same equation as the classical 
ones, namely
\[
:\hat\D_\a:\cdot\G = -\half\DD^\ad :\hat V_{\a\ad}:\cdot\G  
-\dfrac{2}{3}D_\a :\hat S:\cdot\G \ ,
\]
it immediately follows that the same holds for the renormalized $\D^0$, etc.
defined by \equ{q-delta-alpha-0}. It is also evident that
such a renormalization by ``point splitting regularization'' preserves
all the symmetry properties of the corresponding
classical expression. $\D^0$ in particular obeys 
the constraints \equ{d-alpha-constraints}. $V^0$ and $S^0$ are explicitly
$\SS_\G$-variations, the former being a real superfield and the latter a
chiral superfield obeying the constraints \equ{10.8}.

\noindent We can thus write
\eq
\D_\a\cdot\G = \D^0_\a\cdot\G + \D'_\a\cdot\G\ ,
\eqn{def-delta'}
where $\D'_\a$ obeys homogeneous constraints, namely the constraints
\equ{d-alpha-constraints} with the right-hand sides replaced by zero.
The general solution for $\D'_\a$ has the desired form:
\eq
\D_\a'\cdot\G  = -\half\DD^\ad V'_{\a\ad}\cdot\G  
  -\dfrac{2}{3}D_\a S\cdot\G \ ,\quad
\SS_\G V'_{\a\ad}\cdot\G = 0\ ,\quad \SS_\G S\cdot\G  = 0\ ,
\eqn{sol-delta'}
where $V'_{\a\ad}$ and $S$ are $\SS_\G$-invariant, the former being a real
superfield and the latter a chiral superfield\footnote{The proof,
which is rather lengthy, may be found in pages 255-259 of~\cite{ps-book}.}.
This establishes the existence of a BRS invariant supercurrent 
\eq
V_{\a\ad}\cdot\G = :V^0_{\a\ad}:\cdot\G + V'_{\a\ad}\cdot\G\ ,\qquad
\SS_\G V_{\a\ad}\cdot\G = 0\ ,
\eqn{gen-sol-s-cur}
and of a BRS invariant supertrace anomaly $S$ obeying the supertrace
identities \equ{10.14}.

\newpage
\section{Finite Theories}
Our aim is now to show that, in some circumstances, a
supersymmetric gauge theory may be finite. ``Finiteness'' means here
the vanishing of the Callan-Symanzik $\b$-functions, the anomalous
dimensions $\g$ possibly remaining nonzero. In other
words it means the scale invariance of the physical quantities
(e.g. Green functions of BRS-invariant operators), since the
anomalous dimensions do not touch them.

We shall only give a rough sketch of the construction, and advise the
reader to consult the original 
literature~\cite{ps-fin-ijmp,lps-fin-hpa} for complete proofs. 
The starting point is
the relation between the scale anomaly and the anomaly of the
axial R-current, which follows from the supertrace identities
\equ{10.14}. This anomaly is given essentially by the
$\beta$-functions (see \equ{10.21}), which get
contributions from all orders of perturbation theory.

We want now to use the nonrenormalization theorem of the axial
anomaly, which holds for the supercurrent 
constructed in the preceding section (see App. A of
ref.~\cite{lps-fin-hpa}). 
In order to state this theorem we need to expand the
supercurrent anomaly $S$ in a basis different from the one used
in \equ{10.21}. The new expansion reads
\eq
S \cdot\Gamma = \bar{D}^2 \left[ r K^0 + J^{\rm inv} \right]\cdot
\Gamma \ ,
\eqn{11.1}
where $J^{\rm inv}$ is a BRS invariant real insertion and the insertion
$K^0$ is alike a Chern-Simons form. The latter indeed is a
quantum extension of the expression 
\eq
e^{-\f} D^\a e^\f \bar{D}^2 \left( e^{-\f} D_\a e^\f \right) \ ,
\eqn{11.2}
whose real part builds up the super-Yang-Mills action,
and whose imaginary part is a supersymmetric generalization of
the Bardeen current $K^\m$ -- defined by
$\pa_\m K^\m = F_{\m\n} \tilde{F}^{\m \n}$.
More precisely, $K^0$ belongs to a set of insertions $K^q$ obeying ``descent equations''
\eq\ba{l}
\SS_{{\Gamma}} [K^0\cdot \Gamma] =  \bar{D}_{\dot \a} [K^{1
\dot \a}\cdot \Gamma ] \ ,    \es
\SS_{{\Gamma}} [K^{1\dot\a}\cdot \Gamma]  =  \left(
\bar{D}^{\dot \a} D^\a + 2 D^\a \bar{D}^{\dot a}\right) [K^2_\a
. \Gamma ] \ ,    \es
\SS_{{\Gamma}} [K^2_\a\cdot \Gamma]  =  D_\a [K^3\cdot \Gamma ] \ ,
   \es
\SS_{{\Gamma}} [K^3\cdot \Gamma]  =  0 , \qquad \bar{D}_{\dot \a}
[K^3\cdot \Gamma ]  = 0 \ , 
\ea\eqn{11.4}
analogous to the usual ones (see~\cite{pigsor-book}, e.g.), 
where $\SS_{{\Gamma}}$ is the
linearized Slavnov-Taylor operator defined by \equ{lin-slavnov}.
It follows from the first of Eqs. \equ{11.4} that 
$\bar{D}^2[K^0\cdot\G]$ is a BRS invariant insertion.

We don't give here the construction of the insertions $K^q$, and
only state that the last one, $K^3$, is a quantum extension of
\eq
K^3_{\rm class} = \dfrac{1}{3} \tr c^3_+ \ .
\eqn{11.5}
It can be shown\footnote{In~\cite{ps-fin-ijmp,lps-fin-hpa}, 
this result was obtained as
a consequence of the nonrenormalization theorem of chiral
insertions. But the latter holds only in the case of exact
supersymmetry. The argument mentioned presently is more 
general~\cite{pig-sor-antghost}.}~\cite{pig-sor-antghost}, 
that the antighost
equation \equ{antig-eq} forbids any counterterm for the insertion 
$\tr c^3_+$. Hence the latter is UV-finite, and the quantum 
insertion $K^3$ is thus unambiguosly fixed. It follows
then, by solving the descent equations up from the bottom, that
$K^0$ is uniquely defined modulo a BRS invariant
insertion -- absorbed in $J^{\rm inv}$, in Eq. \equ{11.1}, and modulo
a total derivative $\bar{D}$ -- which we shall neglect since we
are interested in $\bar{D}^2 K^0$. 
These remarks are at the basis of the following statement:

\noindent{\it Supersymmetric Nonrenormalization Theorem}.
The coefficient $r$
in \equ{11.1} gets contributions only from one-loop graphs~\cite{lps-fin-hpa}.

The next step consists in comparing both expansions \equ{10.21} and
\equ{11.1} for the supercurrent anomaly. But we need to
modify slightly the former expansion. Choosing
\eq
L^j_i \G = \left( A^j \dfrac{\delta}{\delta A^i} - A^*_i \dfrac{\delta}
{\delta A^*_j} \right) \Gamma \ , 
\eqn{11.6}
in accordance with \equ{10.22} and \equ{10.23}, we split the set
$\{L^j_i\}$ in two subsets
\eq 
\{ L_{0 a} = e_{ aj}^i L^j_i\ , \ a=1,\ 2,\ \cdots \}, \ \{
\L_{1K} = f_{Kj}^i  L^j_i \ , \ K = 1,\ 2, \ \cdots \}
\eqn{11.7}
defined as following. The associated counting operators
\eq
\NN_{a} = e_{aj}^i \NN^j_i
\eqn{11.8 }
form a basis for the counting operators which annihilate the
matter field self-interactions, i.e. the superpotential \equ{matter-s-pot}:
\eq
\NN_a W(A) = 0 \ .
\eqn{inv-s-pot}
The latter conditions  
correspond in fact to the set of renormalizable chiral symmetry
Ward identities
\eq
W_{a} \Gamma = e_{aj}^i \left( \dint d S L^j_i - \dint d \bar{S}
\bar{L}^j_i \right)\G = 0 \ , 
\eqn{11.9}
which constrain these self-interactions. The $L_{1K}$
complete the basis of the linear space spansed by the $L^j_i$.
Moreover the $L_{1K}$ form a basis for the insertion which are
{\em genuinely chiral}, i.e. which are chiral but are not of the
form $\bar{D}^2 (\cdots )$\footnote{This is obvious,
in the classical limit, from the fact that the terms $\int dS\,L_{1K}W(A)$ 
are linearly independent and that $W(A)$ is the only genuinely chiral piece 
of the classical action.}.

\noindent With this, \equ{10.21} becomes
\eq
S = \b_g L_g + \dsum{ijk}{} \b_{ijk} L^{ijk} - \gamma_\f L_\f
- \dsum{a}{} \g_{0a} L_{0a} - \dsum{K}{} \gamma_{1K} L_{1K} -
\dsum{k}{} \gamma_k L_k \ .
\eqn{11.10}
Each of the chiral insertions $L_I$ also possesses 
an expansion of the type \equ{11.1}:
\eq\ba{l}
L_g  =  \bar{D}^2 \left[ \left( \dfrac{1}{128 g^3} + r_g \right)
K^0 + J^{\rm inv}_g \right] + L^c_g \ ,    \es
L_A  =  \bar{D}^2 \left[ r_A K^0 + J^{\rm inv}_A  \right] 
                                        + L^c_A \ ,    \quad
A  = \, (ijk) ,\ \f ,\ 0a,\ 1K,\ k\ , 
\ea\eqn{11.11}
where $r_g$ and the $r_A$ are of order $\hbar$ at least, and
$L^c_g$, $L^c_A$ are genuinely chiral insertions.

\noindent One notes that
\eq
r_\f = 0 \ , \quad r_{1K}= 0 \ . 
\eqn{11.12}
The first equality is due to the possibility of preserving to
all orders~\cite{ps-fin-ijmp} the property
\eq
L_\f = \bar{D}^2 \LL_\f \ ,
\eqn{11.13}
where $\LL_\f$ is BRS-invariant and real. The second one is
obvious from the very definition of $L_{1K}$ as a genuinely
chiral insertion (which also implies $J_{1K}^{\rm inv}=0$).

 We now substitute $S$ and 
the basis elements $L_I$ in \equ{11.10} by
their expressions \equ{11.1} and \equ{11.11}, respectively. Identifying
the coefficient of the resulting $\bar{D}^2 K^0$ term yields the
equation
\eq
r = \b_g \left( \dfrac{1}{128 g^2} + r_g \right) + \b_{ijk} r^{ijk}
 - \g_{0a}r_{0a} - \g_kr_k  \ . 
\eqn{relation-r-beta}
The supersymmetric version of the nonrenormalization theorem
for the axial anomalies also holds for the coefficients $r_{0
a}$ in \equ{11.11} as it did for the coefficient $r$ in \equ{11.1}: the
$r_{0a}$ are exactly given by one-loop graphs. They are
indeed the coefficients of the axial anomalies which break the
conservation of the currents associated to the chiral
invariances\footnote{Eq. \equ{11.11} for $A=0a$ represents the anomalous
conservation laws for the axial Noether currents $J^\m_a$ associated
to the chiral invariances \equ{11.9}. More precisely, $J^\m_a$ is the
$\th\s_\m\tb$ - component of the superfield $J^{\rm inv}_{0a}$, the 
anomalous conservation law is the $\th^2$ - component of 
\equ{11.11} for $A=\,0a\,$. The anomaly is contained in 
the term $\DB^2(r_{0a}K^0)$.}
 \equ{11.9}. From now on we shall restrict ourselves to the
theories for which all axial anomalies vanish:
\eq
r = 0 \ , \quad  r_{0a} = 0 \ .
\eqn{11.14}
This can be achieved by a suitable choice of the representation
in which the matter fields live -- this choice must of course
also assure the vanishing of the gauge anomaly, which is also a 1-loop
problem (see the first remark at the end of Subsection~\ref{ren-de-brs}).

\noindent It can moreover be shown~\cite{lps-fin-hpa} 
that the second of Eqs. \equ{11.14} implies
\eq
r_k = 0 \ .
\eqn{11.15}
 Due to \equ{11.12}, \equ{11.14} and \equ{11.15}, the equation
\equ{relation-r-beta} becomes homogeneous in the $\b$-functions:
\eq
\b_g \left( \dfrac{1}{128 g^2} + r_g \right) + \b_{ijk} r^{ijk} =
0 \ .
\eqn{11.16}
One sees that the first of Eqs. \equ{11.14} corresponds to the vanishing of
$\b_g$ at the one-loop order. Demanding the vanishing of the
$\b_{ijk}$ at this order implies that the coupling constants
$\lambda_{ijk}$ have to be functions of the gauge coupling $g$
(see~\cite{43}). We are therefore motivated 
to demand such a dependence to
all orders. In order to be consistent this dependence must be
given by functions $\lambda_{i j k} (g)$ which are solutions of the
``reduction equations''~\cite{44}
\eq
\b_{ijk} = \b_g \dfrac{d}{dg} \lambda_{ijk} \ . 
\eqn{11.17}
The perturbative existence of a solution to \equ{11.17}
is assured if the
solution at lowest order is ``isolated'', i.e. does not belong
to a continuous family of solutions~\cite{45}. At this stage the
theory depends of the single coupling constants $g$.

The substitution of \equ{11.17} into \equ{11.16} yields
\eq
\b_g \left( \dfrac{1}{128 g^3} + r_g + r_{ijk} \dfrac{d}{dg}
\lambda^{ijk} (g) \right) = \b_g \left( \dfrac{1}{128 g^3} + 
  O(\h)\right) = 0 \ ,
\eqn{11.18}
which implies the vanishing of the $\b$-function:
\eq
\b_g = 0 \ .
\eqn{11.19}
The resulting theory is thus ``finite'', the only ``infinite''
renormalizations being those of the field amplitudes,
characterized by the anomalous dimensions, which may or may not vanish,
but do not correspond to observables anyhow.
\newpage
\appendix \section*{APPENDICES}
\section{Notations and Conventions}\label{notations}
The notations and conventions are those of~\cite{ps-book}.
\subsection{Weyl Spinors and Pauli Matrices}\label{spineurs de Weyl}
\point{Units:} $\h=c=1$
\point{Space-time metric:} $(g_{\m\n}) = \mbox{diag}(1,-1,-1,-1)
    \ ,\quad(\m,\n,\cdots=0,1,2,3)$
\point{Fourier transform:}
\[
f(x) = \dfrac{1}{2\pi}\dint dk\;e^{ikx}\tilde f(k)\ ,\quad
\tilde f(k) = \dint dx\;e^{-ikx}f(k) \ ,\qquad
  \lp\, \pa_\m \leftrightarrow ik_\m\,\rp\ .
\]
\point{Weyl spinor:} $(\p_\a\ ,\ \a =1,2)\ 
   \in\, \mbox{repr. }(\half,0)$ du groupe de Lorentz. \\
  The spinor components are Grassmann variables: $\p_\a\p'_\b=-\p'_\b\p_\a$
\point{Complex conjugate spinor:} 
    $(\bar\p_\ad=(\p_\a)^* ,\ \ad =1,2)\ 
                           \in\, \mbox{repr. }(0,\half)$
\point{raising and lowering of spinor indices:}
  \[\ba{l}
  \p^\a=\e^{\a\b}\p_\b\ ,\quad \p_\a=\e_{\a\b}\p^\b\ ,\es
  \mbox{with }   \e^{\a\b}=-\e^{\b\a}\ ,\quad\e^{12}=1\ ,\quad
  \e_{\a\b}=-\e^{a\b}\ ,\quad \e^{\a\b}\e_{\b\g}=\d^\a_\g\ ,\es
  \mbox{(the same for dotted indices).}
  \ea\]
\point{Derivative with respect to a spinor component:}
  \[\ba{l}
  \dpad{}{\p^\a}\p^\b=\d^{\b}_{\a}\ ,\quad 
     \dpad{}{\p_\a} = \e^{\a\b}\dpad{}{\p^\b}\ ,\es
  \mbox{(the same for dotted indices)}
  \ea\]
\point{Pauli matrices:}
  \[\ba{l}
  \lp\s^\m_{\a\bd}\rp=
  \lp\, \s^0_{\a\bd},\, \s^1_{\a\bd},\, 
              \s^2_{\a\bd},\, \s^3_{\a\bd}\, \rp \es
  \sb_\m^{\ad\b}=\s_\m^{\b\ad}
    =\e^{\b\a}\e^{\ad\bd}\s_{\m\,\a\bd}\ ,\\[4mm]
   {(\s^{\m\n})_\a}^\b=
    \dfrac{i}{2}{\lc \s^\m\sb^\n-\s^\n\sb^\m \rc_\a}^\b   \ , \quad
  {(\sb^{\m\n})^\ad}_\bd=
    \dfrac{i}{2}{\lc \sb^\m\s^\n-\sb^\n\s^\m \rc^\ad}_\bd   \ , 
\ea\]
with
  \[\ba{l}
  \s^0=\lp\matrix{1&0\\0&1}\rp\ ,\quad  \s^1=\lp\matrix{0&1\\1&0}\rp\ ,\quad
    \s^2=\lp\matrix{0&-i\\i&0}\rp\ ,\quad   
                \s^3=\lp\matrix{1&0\\0&-1}\rp\ ,\\[5mm]
  \sb^0 =\s^0\ ,\quad \sb^i=-\s^i=\s_i\ ,
  \s^{0i} = -\sb^{0i} = -i\s^i\ ,\quad 
   \s^{ij} = \sb^{ij} = \e^{ijk}\s^k\ ,\es
   \quad i,j,k=1,2,3\ .
  \ea\]
\point{Summation conventions and complex conjugation:}
Let $\p$ and $\chi$ be two Weyl spinors.
  \[\ba{l}
  \p\chi=\p^\a\chi_\a = -\chi_\a\p^\a = \chi^\a\p_\a = \chi\p \ ,\es
  \psb\chib=\psb_\ad\chib^\ad = -\chib^\ad\psb_\ad 
        = \chib_\ad\psb^\ad  = \chib\psb \ ,\es
  \p\s^\m\chib = \p^\a\smuaad\chib^\ad\ ,\quad  
        \psb\sb_\m\chi = \psb_\ad\sbmuaad\chi_\a\ ,\es
  (\p\chi)^* = \chib\psb = \psb\chib \ ,\es
  (\p\s^\m\chib)^* = \chi\s^\m\psb = -\psb\sb^\m\chi\ ,\es
  (\p\s^{\m\n}\chi)^* = \chib\sb^{\m\n}\psb\ .
\ea\]
\point{Infinitesimal Lorentz transformations of 
         the Weyl spinors:}
  \[\ba{l}
  \mbox{if}\quad  
  \d^{\rm L} x^\m = {\om^\m}_\n x^\n\ ,\quad
     \mbox{with }\om^{\m\n} = -\om^{\n\m}\qquad
      \lp\om^{\m\n}=g^{\n\r}{\om^\m}_\r\rp \ ,\quad\mbox{then:} 
  \es
  \d^{\rm L}\p_\a(x) = 
    \half\om^{\m\n}\lp (x_\m\pa_\n - x_\n\pa_\m)\p_\a(x)
      -\dfrac{i}{2}{(\s_{\m\n})_\a}^\b \p_\b \rp\ ,\es
  \d^{\rm L}\psb^\ad(x) = 
    \half\om^{\m\n}\lp (x_\m\pa_\n - x_\n\pa_\m)\psb^\ad(x)
      +\dfrac{i}{2}{(\sb_{\m\n})^\ad}_\bd \psb^\bd \rp\ .
  
\ea\]

\subsection{Superfields}
 
\point{Superspace:} may be defined as a
``space'' whose ``points'' are characterized by 
even and odd coordinates
\[
\lac (x^\m,\th^\a,\tb^\ad),\ \mu=0,\cdots3,\ \a=1,2,
   \ \ad=\dot 1,\dot 2\rac
\]
where the odd coordinates $\th$  are anticommuting constant Weyl spinors.
\point{Superfields:}
A (classical) superfield is a function in superspace $\f(x,\th,\tb)$,
transforming under infinitesimal translations $P_\m$ and supersymmetry
transformations $Q_a$, $\bar Q_\ad$ with the differential operators
defined by:
\eq\ba{l}
\d^P_\m \f = \pa_\m\f\ ,\es
\d^Q_\a \f = \lp \dpad{}{\th^\a} + i\smuaad\tb^\ad\pa_\m\rp \f , \es
\d^{\bar Q}_\ad \f = \lp -\dpad{}{\tb^\ad} - i\th^\a\smuaad\pa_\m\rp \f\ ,
\ea\eqn{def-delta-susy}
obeying the algebra
\eq\ba{l}
\lac \d^Q_\a,\d^{\bar Q}_\ad \rac = -2i\smuaad\d^P_\m\ ,\es
\mbox{(the other (anti)commutators vanishing)}\ .
\ea\eqn{alg-delta-susy}
Due to the anticommutivity of the $\th_\a$ and $\tb_\ad$, one can write
a finite expansion for the superfield $\f$ 
(we take it real, $\bar\f=\f$):
\eq\ba{l}
\f(x,\th,\tb) = C(x) + \th\chi(x) + \tb\bar\chi(x) + \half\th^2M(x)
+ \half\tb^2\bar M(x)  \es\phantom{\f(x,\th,\tb) =}
+ \th\s^\m\tb v_\m(x) + \half\tb^2\th\la(x)
+ \half\th^2\tb\bar\la + \frac 1 4  \th^2\tb^2D(x)\ ,
\ea\eqn{theta-exp}
where the components are ordinary space-time fields.
\point{Covariant derivatives}
They are defined such as to anticommute with the supersymmetry
transformation rules \equ{def-delta-susy}:
\eq
D_\a  =  \dpad{}{\th^\a} - i\smuaad\tb^\ad\pa_\m \ ,\quad
\DB_\ad =  -\dpad{}{\tb^\ad} + i\th^\a\smuaad\pa_\m \ .
\eqn{def-cov-der}
They obey the algebra
\eq\ba{l}
\lac D_\a,\DB_\ad \rac = 2i\smuaad\d^P_\m\ ,\es
\mbox{(the other anticommutators vanishing)}\ .
\ea\eqn{alg-cov-der}
\point{Chiral superfields:}
A chiral superfield $A$ is defined through the constraint
\[
\DB_\ad A(x,\th,\tb) = 0\ ,
\]
The complex conjugate constraint defines an antichiral superfield 
$\bar A$: 
\[
D_\a \bar A(x,\th,\tb) = 0\ .
\]
These constraints can be solved {\it algebraically}, 
with the help of the
commutation rules \equ{com-chiral}. The result is the following,
expanded in component fields:
\eq\ba{l}
A(x,\th,\tb) = e^{-i\th\s^\m\tb\pa_\m}A_{\rm chiral}(x,\th)=
  e^{-i\th\s^\m\tb\pa_\m}\lp A(x) + \th\p(x) + \th^2 F(x)\rp\ ,\es
\bar A(x,\th,\tb) = e^{i\th\s^\m\tb\pa_\m}
               \bar A_{\rm antichiral}(x,\tb)=
  e^{i\th\s^\m\tb\pa_\m}\lp \bar A(x) + \tb\psb(x) + \tb^2 \bar F(x)\rp\ .
\ea\eqn{comp-chiral}
Note that the same symbol $A$ is used for both the chiral superfield
$A(x,\th,\tb)$ and its $\th=0$ component $A(x)$.
\point{Chiral and antichiral representations:}
It is possible to perform changes of superspace coordinates 
in such a way that the covariant derivatives, either $\DB$ or $D$, take
a simple form. This leads to the two following representations for the
superfields.
\begin{enumerate} 
\item[1.] The chiral representation, defined by
\eq
\f_{\rm (chir\ rep)}(x,\th,\tb) = \f(x+i\th\s\tb,\th,\tb)\ .
\eqn{chir-rep}
In this representation, the transformation laws \equ{def-delta-susy}
and the covariant derivatives \equ{def-cov-der} take the form
\eq\ba{ll}
\d^Q_\a  = \dpad{}{\th^\a}\ ,\qquad &
\d^{\bar Q}_\ad \f =  -\dpad{}{\tb^\ad} - 2i\th^\a\smuaad\pa_\m
              \ ,\es 
D_\a  =  \dpad{}{\th^\a} - 2i\smuaad\tb^\ad\pa_\m \ ,\qquad&
\DB_\ad =  -\dpad{}{\tb^\ad} \ .
\ea\eqn{chir-rep-op}
In the chiral representation, a chiral superfield is independent of
$\tb$: its form is given by the first of Eqs.~\equ{comp-chiral} 
without the exponential factor.
\item[2.] The antichiral representation, defined by
\eq
\f_{\rm (antichir\ rep)}(x,\th,\tb) = \f(x-i\th\s\tb,\th,\tb)\ .
\eqn{antichir-rep}
In this representation, the transformation laws \equ{def-delta-susy} and
the covariant derivatives \equ{def-cov-der} take the form
\eq\ba{ll}
\d^Q_\a \f =  \dpad{}{\th^\a} + 2i\smuaad\tb^\ad\pa_\m\ , \qquad&
\d^{\bar Q}_\ad \f = -\dpad{}{\tb^\ad} \ ,
              \ ,\es 
D_\a  =  \dpad{}{\th^\a}  \ ,\qquad&
\DB_\ad =  -\dpad{}{\tb^\ad} + 2i\th^\a\smuaad\pa_\m \ .
\ea\eqn{antichir-rep-op}
In the antichiral representation, an
antichiral superfield is independent of
$\th$: its form is given by the second of Eqs.~\equ{comp-chiral} 
without the exponential factor.
\end{enumerate}
\point{``Tensor calculus'':}
Products of superfields are superfields. \\
Products of chiral superfields of the same chirality are chiral.\\
The double derivative $\DB^2\f$ of a superfield is a chiral superfield.
\point{Superspace integration:}
The integral with respect of a Grassmann variable $\th$
being defined~\cite{berezin}
 by the derivative $\pa/\pa\th$, one defines the integral
of a superfield $\f$, or of a (anti)chiral superfield 
$A$ ($\bar A$) by
\eq
\iv\f=\dint d^4x\, D^2\DB^2\f\ ,\quad
\is  A=\dint   d^4x\, D^2 A\ ,\quad
\isb \bar A=\dint d^4x\, \DB^2\bar A\ .
\eqn{integr}
The usual formula for the integration by part holds -- but only for the
full superspace measure $dV$ -- since
\eq
\iv D_\a\f=0\ ,
\eqn{part-int}
provided $\f(x,\th,\tb)$ decreases sufficiently rapidly at infinity in
$x$-space. It also follows from the latter equation
 that the integrals \equ{integr} are {\it invariant}
under the supersymmetry transformations.
 
\point{Superspace Dirac distributions:}
We use the notation:
\[\ba{l}
F(1,2,\cdots) = F(x_1,\th_1,\tb_1,x_2,\th_2,\tb_2,\cdots)
\ ,\quad
\th_{12}=\th_1-\th_2 \ .
\ea\]
The delta functions are defined by
\eq\ba{l}
\dint dV(2)\d_V(1,2)\f(2) = \f(1)\es
\dint dS(2)\d_S(1,2)A(2) = A(1)\es
\dint d\SB(2)\d_{\bar S}(1,2)\bar A(2) = \bar A(1)\ ,
\ea\eqn{def-delta-funct}
and are expressed by
\eq\ba{l}
\d_V(1,2) = \dfrac{1}{16}\th_{12}^2\tb_{12}^2\,\d^4(x_1-x_2)\es
\d_S(1,2) = -e^{i(\th_1\s\tb_2-\th_2\s\tb_1)\pa}
               \,\dfrac{1}{4}\th_{12}^2\d^4(x_1-x_2)\es
\d_\SB(1,2) = -e^{i(\th_1\s\tb_2-\th_2\s\tb_1)\pa}
                 \,\dfrac{1}{4}\tb_{12}^2\d^4(x_1-x_2)\ .\es
\ea\eqn{delta-funct}
One has:
\eq\ba{l}
\d_S(1,2) = \DB^2\d_V(1,2)\ ,\quad \d_\SB(1,2) = D^2\d_V(1,2)\ ,\es
\DB^2\d_\SB(1,2)=e^{i(\th_1\s\tb_2-\th_2\s\tb_1-\th_{12}\s\tb_{12})\pa}
         \d^4(x_1-x_2)\ ,\es
D^2\d_S(1,2)    =e^{i(\th_1\s\tb_2-\th_2\s\tb_1+\th_{12}\s\tb_{12})\pa}
         \d^4(x_1-x_2)\ .
\ea\eqn{dd-delta}
\point{Functional differentiation:}
\eq
\dfud{\f(1)}{\f(2)}=\d_V(1,2)\ ,\quad\dfud{A(1)}{A(2)}=\d_S(1,2)\ ,
\quad \dfud{\AB(1)}{\AB(2)}=\d_\SB(1,2)\ .
\eqn{funct-diff}

\point{Supersymmetry transformations of the components:}
The components of a superfield, defined by the expansion 
\equ{theta-exp}, transform under supersymmetry as
\eq\ba{ll}
\d_\a C = \chi   &\bar\d_\ad C = \bar\chi \es
\d_\a \chi^\b = \d_\a^\b M   
   &\bar\d_\ad {\bar\chi}^\bd = -\d_\ad^\bd \bar M\es
\d_\a\bar\chi_\ad = \smuaad (v_\m+i\pa_\m C)
   &\bar\d_\ad\chi_\a = -\smuaad (v_\m-i\pa_\m C) \es
\d_\a M = 0  &\bar\d_\ad \bar M = 0  \es
\d_\a \bar M = \la_\a-i(\s^\m\pa_\m\bar\chi)_\a 
   &\bar\d_\ad  M = \bar\la_\ad+i(\pa_\m\chi\s^\m)_\ad \es
\d_\a v_\m = \half(\s_\m\bar\la)_\a 
           - \frac{i}{2}(\s^\n\sb_\m\pa_\n\chi)_\a
 \quad  &\bar\d_\ad v_\m = \half(\la\s_\m)_\ad 
           + \frac{i}{2}(\pa_\n\bar\chi\sb_\m\s^\n)_\ad \es
\d_\a\la^\b = \d_\a^\b D + i(\s^\n\sb^\m)_\a{}^\b\pa_\n v_\m
   &\bar\d_\ad{\bar\la}^\bd = -\d_\ad^\bd D 
     + i(\sb^\m\s^\n)^\bd{}_\ad \pa_\n v_\m \es
\d_\a\bar\la_\ad = i\smuaad\pa_\m M  
   &\bar\d_\ad\la_\a = i\smuaad\pa_\m\bar M  \es
\d_\a D = -i(\s^\m\pa_\m\bar\la)_\a
   &\bar\d_\ad D = i(\pa_\m\la\s^\m)_\ad
\ea\eqn{transf-fi}
For the components of the chiral and antichiral superfields 
\equ{comp-chiral}, one has
\eq\ba{ll}
\d_\a A = \p_\a   &\db_\ad \bar A = \psb_\ad   \es
\d_\a\p^\b = 2\d_\a^\b F   &\db_\ad\psb^\bd = -2\d_\ad^\bd\bar F \es
\d_\a F = 0   &\db_\ad\bar F = 0  \es
\d_\a\bar A = 0   &\db_\ad A = 0  \es
\d_\a\psb_\ad = 2i\smuaad\pa_\m\bar A \quad
   &\db_\ad\p_\a = 2i\smuaad\pa_\m A  \es
\d_\a\bar F = -i(\s^\m\pa_\m\psb)_\a \quad
   &\db_\ad F = i(\pa_\m\p\s^\m)_\ad 
\ea\eqn{transf-chi}
\subsection{Some Useful Formula}
\point{Algebra of Pauli matrices:}
\eq\ba{l}
\e_{\a\b}\s^\m_{\g\gd} + \e_{\b\g}\s^\m_{\a\gd} + 
  \e_{\g\a}\s^\m_{\b\gd} = 0\ ,\es
\e_{\a\b}\e_{\g\d} + \e_{\b\g}\e_{\a\d} + 
  \e_{\g\a}\e_{\b\d} = 0\ .
\ea\eqn{fierz}

\eq\ba{ll}
\s^\m_{\a\ad}\s_\m^{\b\bd} = 2 \d^\b_\a\d^\bd_\ad\ ,
\quad &\s^\m_{\a\ad}\s_\m{}_{\b\bd} = 2 \e_{\a\b}\e_{\ad\bd}  \ ,\es
\s_\m{}_{\a\ad}\s^{\m\n}_{\b\g}
 =i(\e_{\a\b}\s^\n_{\g\ad}+\e_{\a\g}\s^\n_{\b\ad})\ ,\quad
&\s_\m{}_{\a\ad}\sb^{\m\n}_{\bd\gd}
 =-i(\e_{\ad\bd}\s^\n_{\a\gd}+\e_{\ad\gd}\s^\n_{\a\bd})\ ,\es
\s_{\m\n}^{\a\b}\s^{\m\n}_{\g\d}=-4(\d^\a_\g\d^\b_\d+\d^\a_\d\d^\b_\g)
\ ,\quad
&\sb_{\m\n}^{\ad\bd}\sb^{\m\n}_{\gd\dd}=
  -4(\d^\ad_\gd\d^\bd_\dd+\d^\ad_\dd\d^\bd_\gd)\ ,\es
\sb_{\m\n}^{\ad\bd}\s^{\m\n}_{\g\d}=0\ ,&\es
\ea\eqn{smn-smn}

\eq\ba{l}
\half \e^{\m\n\r\s}\s_{\r\s}= -i \s^{\m\n}\ ,\quad
\half \e^{\m\n\r\s}\sb_{\r\s}= i \sb^{\m\n}\ ,\es
\e_{\m\n\r}{}^\tau\s_{\tau\la} =  
 i\s_{\m\n}g_{\r\l} - i\s_{\m\r}g_{\n\l} + i\s_{\n\r}g_{\m\l} \ ,\es
\e_{\m\n\r}{}^\tau\sb_{\tau\la} =  
 -i\sb_{\m\n}g_{\r\l} + i\sb_{\m\r}g_{\n\l} - i\sb_{\n\r}g_{\m\l} \ ,
\ea\eqn{duality}
\[\ba{l}
\mbox{with:}\quad \e_{0123} = 1 = -\e^{0123}\ ,\es
\phantom{\mbox{with:}\quad}
\half\e^{\m\n\r\s}\e_{\r\s\la\tau} =
        -(\d^\m_\la\d^\n_\tau-\d^\n_\la\d^\m_\tau)  \ .
\ea\]

\eq
\s_\m\sb_\n = g_{\m\n}-i\s_{\m\n}\ ,\quad 
\sb_\m\s_\n = g_{\m\n}-i\sb_{\m\n}\ .
\eqn{s-sbar}

\eq\ba{l}
\s^\m\sb^\n\s^\r = g^{\m\n}\s^\r + g^{\n\r}\s^\m - g^{\m\r}\s^\n
  - i\e^{\m\n\r\la}\s_\la  \ ,\es
\sb^\m\s^\n\sb^\r = g^{\m\n}\sb^\r + g^{\n\r}\sb^\m - g^{\m\r}\sb^\n
  + i\e^{\m\n\r\la}\sb_\la  \ .
\ea\eqn{s-s-s}

\eq\ba{l}
\s^{\m\n}\s^{\r} = i\s^\m g^{\n\r} - i\s^\n g^{\m\r} 
     + \e^{\m\n\r\la}\s_\la\ ,\es
\s^{\r}\sb^{\m\n} = i\s^\n g^{\r\m} - i\s^\m g^{\r\n} 
     + \e^{\m\n\r\la}\s_\la\ ,\es
\sb^{\m\n}\sb^{\r} = i\sb^\m g^{\n\r} - i\sb^\n g^{\m\r} 
     - \e^{\m\n\r\la}\sb_\la\ ,\es
\sb^{\r}\s^{\m\n} = i\sb^\n g^{\r\m} - i\s^\m g^{\r\n} 
     - \e^{\m\n\r\la}\sb_\la\ .
\ea\eqn{smn-sr}

\eq\ba{ll}
\s_\m\sb_\n\s^\m = -2\s_\n\ , &\sb_\m\s_\n\sb^\m = -2\sb_\n\ ,\es
\s_\m\sb_{\r\la}\sb^\m = \sb_\m\s_{\r\la}\s^\m = 0\ ,\quad
    &\s^{\m\n}\s_{\r\la}\s_{\m\n} = -4\s_{\r\la}\ ,\es
\s_\m\sb^{\m\n} = 3i\s^\n\ , &\sb_\m\s^{\m\n} = 3i\sb^\n\ ,\es
\sb^{\m\n}\sb_\n = 3i\sb^\m\ ,&\s^{\m\n}\s_\n = 3i\s^\m\ ,\es
\s_{\m\n}\s^{\m\n} = 12\ .
\ea\eqn{contractions}

\eq\ba{l}
\s^{\m\n}\s^{\r\la} = g^{\m\r}g^{\n\la}-g^{\m\la}g^{\n\r}+i\e^{\m\n\r\la}
  +i\s^{\m\la}g^{\n\r}-i\s^{\n\la}g^{\m\r}
      +i\s^{\n\r}g^{\m\la}-i\s^{\m\r}g^{\n\la}\ ,\es
\sb^{\m\n}\sb^{\r\la} = g^{\m\r}g^{\n\la}-g^{\m\la}g^{\n\r}-i\e^{\m\n\r\la}
  +i\sb^{\m\la}g^{\n\r}-i\sb^{\n\la}g^{\m\r}
      +i\sb^{\n\r}g^{\m\la}-i\sb^{\m\r}g^{\n\la}\ ,\es
\{\s^{\m\n},\s^{\r\la}\} = \tr\s^{\m\n}\s^{\r\la} =
  2(g^{\m\r}g^{\n\la}-g^{\m\la}g^{\n\r}+i\e^{\m\n\r\la})\ ,\es
\{\sb^{\m\n},\sb^{\r\la}\} = \tr\sb^{\m\n}\sb^{\r\la} =
  2(g^{\m\r}g^{\n\la}-g^{\m\la}g^{\n\r}-i\e^{\m\n\r\la})\ ,\es
[\s^{\m\n},\s^{\r\la}] = 2i(\s^{\m\la}g^{\n\r}-\s^{\n\la}g^{\m\r}
      +\s^{\n\r}g^{\m\la}-\s^{\m\r}g^{\n\la})\ ,\es
[\sb^{\m\n},\sb^{\r\la}] = 2i(\sb^{\m\la}g^{\n\r}-\sb^{\n\la}g^{\m\r}
      +\sb^{\n\r}g^{\m\la}-\sb^{\m\r}g^{\n\la})\ 
\ea\eqn{smn-srl}

\point{Algebra of covariant derivatives:}
\eq\ba{l}
D_\a\lp e^{i\th\s^\m\tb\pa_\m} \f \rp =
  e^{i\th\s^\m\tb\pa_\m} \dpad{}{\th^\a}\f\ ,\es
\DB_\ad\lp e^{-i\th\s^\m\tb\pa_\m} \f \rp =
  e^{-i\th\s^\m\tb\pa_\m} \lp-\dpad{}{\tb^\ad}\rp\f\ .
\ea\eqn{com-chiral}

\eq\ba{l}
[D_\a,\DB^2]=4i(\s^\m\DB)_\a\pa_\m\ ,\quad
[\DB_\ad,D^2]=-4i(D\s^\m)_\ad\pa_\m \es
[D^2,\DB^2]=8iD\s^\m\DB\pa_\m+16\pa^2 = 
           -8i\DB\sb^\m D\pa_\m-16\pa^2\es
D\DB^2D=\DB D^2\DB\es
D\DB_\ad D = -\half\DB_\ad D^2-\half  D^2\DB_\ad\ , \quad
\DB D_\a \DB = -\half D_\a \DB^2-\half  \DB^2D_\a
\ea\eqn{com-cov-der}
The following operators are projectors:
\eq\ba{l}
P^{\rm T} = \dfrac{D\DB^2D}{8\pa^2}\ , \quad
P^{\rm L} = -\dfrac{D^2\DB^2+\DB^2D^2}{16\pa^2}  \es
\lp P^{\rm T}\rp^2= P^{\rm T}  \ , \quad
\lp P^{\rm L}\rp^2=P^{\rm L}\ ,\quad P^{\rm T}P^{\rm L}= 0
  \ ,\quad P^{\rm T}+P^{\rm L}= 1\ .
\ea\eqn{projectors}
Applied to the superspace Dirac distribution $\d_V$ \equ{delta-funct} 
they give
\eq\ba{l}
P^{\rm T}\d_V(1,2) = 
\dfrac{1}{8\pa^2}\lp 1+\dfrac{1}{4}\th_{12}^2\tb_{12}^2\pa^2 \rp
e^{i(\th_1\s\tb_2-\th_2\s\tb_1)\pa} \d^4(x_1-x_2)\ ,\es
P^{\rm L}\d_V(1,2) = 
\dfrac{1}{8\pa^2}\lp -1+\dfrac{1}{4}\th_{12}^2\tb_{12}^2\pa^2\rp
e^{i(\th_1\s\tb_2-\th_2\s\tb_1)\pa} \d^4(x_1-x_2) \ .
\ea\eqn{proj-delta-fct}

\newpage
\section{Generating Functionals}\label{fonctionelles}
\newcommand{\ex}{{\rm e}}
The content of this appendix is taken from ref.~\cite{ps-book}.
Let us consider a theory involving a set of fields $\f_i(x)$ in 
$D$--dimensional space--time\footnote{Space--time point coordinates are 
denoted by $(x^\m,\ \m=0,\cdots, D-1)$.},
with the index $i$ denoting the species as well as the
spin and internal degrees of freedom. The (classical) dynamics is
defined by the action
\eq
S(\f) = \dint dx\LL(x)  = S_0(\f) + S_{\rm int}(\f)   \ .
\eqn{class-action}
The Lagrangian has the general form
\eq
\LL(x)= \half \f_i(x) K^{ij}(\pa)\f_j(x)  + \LL_{\rm int}
         =  \LL_0 + \LL_{\rm int}       \ .
\eqn{Lagrangian}
$K^{ij}(\pa)$ is some invertible differential operator, usually
 a polynomial of second order in $\pa$ for the bosonic fields
and of first order for the fermionic
ones: the quadratic piece $\LL_0$ of the Lagrangian corresponds to the free
theory whereas $\LL_{\rm int}$ describes the interactions.
\subsection{The Green Functional}\label{fctelle de green}
The objects of the corresponding quantum theory
one wants to compute are
the Green functions, i.e., the vacuum expectation values of the
time--ordered
products of field operators:
\eq
G_{i_1\cdots i_N}(x_1,\cdots,x_N) =
   \vev{T \f_{i_1}(x_1)\cdots\f_{i_N}(x_N)}\ .
\eqn{green-functions}
These Green functions may be collected together in the generating
functional $Z(J)$, a formal power series in the ``classical sources''
$J^i(x)$\footnote{The Planck constant $\h$ is set equal to 1 in the
main text.}:
\eq
Z(J) = \dsum{N=0}{\infty} \dfrac{(-1/\hbar)^N}{N!}
                           \dint dx_1\cdots dx_N\,
  J^{i_1}(x_1)\cdots J^{i_N}(x_N)\, G_{i_1\cdots i_N}(x_1,\cdots,x_N)\ .
\eqn{green-functional}
The Green functions  are tempered distributions.
The sources $J^i(x)$ thus
belong to the set of Schwartz fast decreasing $C^\infty$ functions
(``test functions'').

The Green functional \equ{green-functional}
is {\it formally} given by the Feynman path
integral
\eq
Z(J) = \NN \dint \DD\f\,
 \ex^{ -\frac{1}{\hbar}\lp S(\f) + \int dx J^i(x)\f_i(x)\rp} \ ,
\eqn{path-integral}
where $\NN$ is some (generally ill--defined) numerical factor.
The solution for the free theory ($\LL_{\rm int}=0$ in \equ{Lagrangian})
is given by
\eq
Z_{\rm free}(J) = \ex^{\,{  \frac{1}{2\hbar^2}\int dx_1dx_2
   J^{i_1}(x_1)J^{i_2}(x_2) \D_{i_1 i_2}(x_1,x_2) } }\ ,
\eqn{free-green}
where $\D_{i_1 i_2}(x_1,x_2)$ is the Stueckelberg--Feynman
free causal propagator, obtained by
inverting\footnote{Signs in the following equations
correspond to the case where the fields $\f_i$ are all bosonic. The
reader may generalize to the case where fermionic fields are also
present.} the wave operator $K^{ij}(\pa)$
of \equ{Lagrangian}:
\eq
K^{ij}\D_{jk}(x,y) = \h \d_k^i \d^D(x-y)\ .
\eqn{prop-equation}

In the case of the full interacting theory a formal solution is given
by~\cite{ramond}
\eq
Z(J) = \NN \ex^{ -\frac{1}{\hbar}
          S_{\rm int}\lp -\hbar\fud{}{J}\rp } Z_{\rm free}(J)\ .
\eqn{pert-green}
This expression leads to the well--known perturbative expansion
of
the Green functions in terms of Feynman graphs.
\subsection{The Connected and the Vertex
 Functionals}\label{fctelle conn et vertex}
Let us introduce two more functionals.
The total contribution of the {\it connected} graphs to a Green function is
called a truncated or a connected Green function. The generating functional
of the connected Green functions
\eq\ba{l}
\ZC(J) = \es
 \quad \dsum{N=1}{\infty} \dfrac{(-1/\hbar)^{N-1}}{N!}
       \dint dx_1\cdots dx_N\,   J^{i_1}(x_1)\cdots J^{i_N}(x_N)\,
        \vev{T \f_{i_1}(x_1)\cdots\f_{i_N}(x_N)}_{\rm conn}
\ea\eqn{connected-functional}
is related to the Green functional $Z(J)$ by
\eq
Z(J) = \ex^{-\frac{1}{\hbar}\ZC(J)}\ .
\eqn{exponentiation}

A {\it one--particle irreducible} (1PI) graph is a connected
graph, amputated from its external legs,
which remains connected after cutting any internal line.
The total contribution of these graphs to an (amputated) connected
Green function is called a
1PI or vertex function. The generating functional of the
vertex functions reads
\eq\ba{l}
\G(\f^{\rm class}) = \dsum{N=2}{\infty} \dfrac{1}{N!}
       \dint dx_1\cdots dx_N\,   \f^{\rm class}_{i_1}(x_1)\cdots
                             \f^{\rm class}_{i_N}(x_N)\,
                     \G^{i_1\cdots i_N}(x_1,\cdots ,x_N)  \es
\   \G^{i_1\cdots i_N}(x_1,\cdots x_N) =
        \vev{T \f_{i_1}(x_1)\cdots\f_{i_N}(x_N)}_{\rm 1PI}\ ,
\ea\eqn{vertex-functional}
 where the arguments $\f^{\rm class}$, the ``classical fields'', are
Schwartz fast decreasing test functions. Later on we shall suppress
the superscript ``class'', no confusion between the classical field and
the corresponding quantum field being expected.
The vertex functional is related to the connected functional by a
Legendre transformation:
\eq
\G(\f) = \ZC(J) - \left. \dint dx J^i(x) \f_i(x)\right|_{
   \f_i(x) = \fud{\ZC}{J^i(x)}   }   \ .
\eqn{legendre}
In the right--hand side, $J(x)$ is replaced by the solution
$J(\f)(x)$ of the  equation 
$$\f_i(x) = \d\ZC/\d J^i(x)\ .$$
The inverse Legendre transformation is given by
\eq
\ZC(J) = \G(\f) + \left. \dint dx J^i(x) \f_i(x)\right|_{
J^i(x) = -\fud{\G}{\f_i(x)} } \ .
\eqn{inverse-legendre}
 We have assumed that
the vacuum expectation values of the field
variables are zero. One has thus
\eq
\left. \dfud{\ZC}{J^i(x)}\right|_{J=0} = 0\ ,\qquad
    \left. \dfud{\G}{\f_i(x)}\right|_{\f=0} =0 \ .
\eqn{zero-vev}
\begin{remark}
For the two--points functions the Legendre transformation yields:
\eq
\dint dy \G^{ij}(x,y)  \vev{T \f_j(y)\f_k(z)}_{\rm conn} =
     -\d^i_k \d^D(x-y)\ ,
\eqn{gz=-1}
which is the all order generalization of \equ{prop-equation}.
\end{remark}
\subsection{Expansion in $\h$}\label{exp hbar}
{}From its definition and the formula \equ{pert-green} for the
perturbative expansion of the Green functional
one can easily check that the vertex functional
can be written as a formal power series in $\hbar$:
\eq
\G(\f) = \dsum{n=0}{\infty} \hbar^n  \G^{(n)}(\f)\ ,
\eqn{hbar-expansion}
the order $n$ corresponding to the contributions of the $n$--loop
graphs.
In order to prove this statement,
let us consider the contribution of
a 1PI diagram  consisting of
$I$ internal lines, $V$ vertices and $L$ loops. Counting a factor $\h$ for
each internal line, a factor $\h^{-1}$ for each vertex and an overall
factor $\h$ due to the factor $\h^{-1}$ in \equ{exponentiation}, we
find the value $I-V+1$ for the total power in $\h$.
The result then follows from the topological identity
\eq
L=I-V+1 \ ,
\eqn{euler-top-id}
due to Euler.
The zeroth order
\eq
\G^{(0)}(\f) = S(\f)
\eqn{order-zero}
is the classical action \equ{class-action}. This is obvious since the
only 1PI zero-loop graphs -- the 1PI tree graphs --
are the trivial ones,  i.e., those containing a single vertex, and this
vertex corresponds to a term of the interaction Lagrangian.
In this approximation the Legendre transform $Z^{c(0)}(J)$ of
$\G^{(0)}(\f)$ generates the
connected Green functions, given by the connected tree Feynman graphs.
$\G^{(n)}$  corresponds to the contributions of the $n$--loop
graphs.
\subsection{Composite Fields}\label{champs comp}
We are also interested in Green functions involving composite field
operators. Such operators appear in particular in theories invariant
under field transformations which depend nonlinearly on the
fields -- e.g. the BRS transformations in (super) Yang-Mills theories. 
Let us thus consider field operators $Q^p(x)$,
corresponding to local field polynomials $Q^p_{\rm class}(x)$
in the classical theory.
If one performs again the construction above, but starting with the new
classical interaction
\eq
S_{\rm int}(\f,\r) = S_{\rm int}(\f)
                + \dint dx \r_p(x)Q^p_{\rm class}(x)
\eqn{rho-action}
depending on the ``external fields'' $\r_p(x)$, one obtains a new Green
functional~\cite{symanzik}
\eq\ba{rl}
Z(J,\r) = \dsum{N=0}{\infty}\dsum{M=0}{\infty}
            \dfrac{(-1/\hbar)^N}{ N!}& \dfrac{(-1/\hbar)^M}{ M!}
                    \dint dx_1\cdots dx_N\, \dint dy_1\cdots dy_M\,\es
&J^{i_1}(x_1)\cdots J^{i_N}(x_N)\,
              \r_{p_1}(y_1)\cdots \r_{p_M}(y_M)\,   \es
&\  \vev{T \f_{i_1}(x_1)\cdots\f_{i_N}(x_N)\,
                  Q^{p_1}(y_1)\cdots Q^{p_M}(y_M)  }\ ,
\ea\eqn{rho-green-functional}
which generates the Green functions with insertions of the local
composite quantum  fields $Q^p(x)$.

The connected functional $\ZC(J,\r)$
and the vertex functional $\G(\f,\r)$ involving these
composite fields are related to $Z(J,\r)$ via the generalizations of
\equ{exponentiation}, \equ{legendre} and \equ{inverse-legendre}:
\eq
Z(J,\r) = \ex^{-\frac{1}{\hbar}\ZC(J,\r)} \ ,
\eqn{rho-exponentiation}
\eq
\G(\f,\r) = \ZC(J,\r) - \left. \dint dx J^i(x) \f_i(x)\right|_{
   \f_i(x)  = \fud{\ZC}{J^i(x)} }   \ ,
\eqn{rho-legendre}
and
\eq
\ZC(J,\r) = \G(\f,\r) + \left. \dint dx J^i(x) \f_i(x)\right|_{
  J^i(x) = -\fud{\G}{\f_i(x)} } \ .
\eqn{rho-inverse-legendre}
In particular
\eq
-\hbar \left.\dfud{Z}{\r_p(y)}\right|_{\r=0}
  := Q^p(y)\cdot Z(J)\ ,
\eqn{insertion}
generates the Green functions
\eq
\vev{T Q_p(y)\, \f_{i_1}(x_1)\cdots\f_{i_N}(x_N)}\ ,
\eqn{inserted-function}
whose Feynman graphs contain a new vertex corresponding to the insertion
of the field polynomial $Q^p_{\rm class}(y)$ (with possible quantum
corrections). In the same way
\eq
\left.\dfud{\ZC}{\r_p(y)}\right|_{\r=0}
  = Q^p(y)\cdot \ZC(J)\ ,
\eqn{connected-insertion}
and
\eq
\left.\dfud{\G}{\r_p(y)}\right|_{\r=0}
  = Q^p(y)\cdot \G(\f)\ ,
\eqn{vertex-insertion}
generate connected and 1PI Green functions, respectively, with the
insertion of the operator $Q^p(y)$. The zeroth order term of
the loop ($\h$) expansion of \equ{vertex-insertion} coincides with the
classical field polynomial which is the starting point of
the perturbative construction of the quantum insertion $Q^p$:
\eq
Q^p(y)\cdot \G(\f) = Q^p_{\rm class}(y) + O(\h)  \ .
\eqn{class-lim-insert}


\end{document}